\documentclass[fp,twocolumn]{jpsj3}
\usepackage{txfonts}
\usepackage{graphicx,bm,color,braket}

\title{Charging in a Superconducting Vortex Due to the Three Force Terms in Augmented Eilenberger Equations}

\author{Hikaru Ueki, Marie Ohuchi, and Takafumi Kita}
\inst{Department of Physics, Hokkaido University, Sapporo 060-0810, Japan} 

\abst{
We derive augmented Eilenberger equations  
that incorporate the following missing force terms: (i) the Lorentz force, 
(ii)  the pair-potential gradient (PPG) force, 
and (iii) the pressure difference arising from the slope in the density of states (DOS). 
Recently, augmented Eilenberger equations with the Lorentz and PPG forces have been derived microscopically by studying the Hall and charging effects in superconductors, 
but the pressure due to the slope in the DOS has not yet been considered in augmented Eilenberger equations, 
despite phenomenological indications that it is a charging mechanism in a vortex of type-II superconductors.
This newly added pressure is called ``the SDOS pressure".  
We calculate the charging in an isolated vortex of an $s$-wave superconductor with a spherical Fermi surface
using the augmented Eilenberger equations incorporating the Lorentz force, PPG force, and SDOS pressure. 
When we compare the charge densities due to the three force terms in the augmented Eilenberger equations, 
the vortex-core charging due to the SDOS pressure is larger than that due to the other forces near the superconducting transition temperature. 
Thus, when we calculate the charging in an isolated vortex of a superconductor with a finite slope in the DOS, 
we should consider not only the Lorentz and PPG forces but also the SDOS pressure. 
}

\begin{document}
\maketitle

\section{Introduction \label{sec:I}}  

The {\em vortex-core charging} in type-II superconductors has been pointed out to 
be related to the sign change of the flux-flow Hall conductivity 
\cite{Iye,Artemenko,Hagen90,Affronte,Hagen91,Zavaritsky,Chien,Luo,Hagen93}, 
and numerous studies on the charging of a superconducting vortex have been carried out 
\cite{Khomskii,Feigel'man,Blatter,Hayashi,Matsumoto,Chen,Machida,Knapp,Kumagai}.
However, the forces responsible for the charging of a superconducting vortex are not fully understood.
This is because 
all the force terms used to describe charging in superconductors are missing from the standard Eilenberger equations 
\cite{Eilenberger}
(i.e., the quasiclassical equations of superconductivity) 
used to study superconductors in a magnetic field microscopically 
\cite{Kramer,Klein,Schopohl,Ichioka96}. 
Although there have been several numerical calculations of the vortex-core charging in superconductors 
\cite{Hayashi,Matsumoto,Chen,Machida,Knapp}
based on the Bogoliubov--de Gennes (BdG) equations \cite{Caroli,deGennes}, 
these equations are not suitable for studying the charging mechanism.
The purpose of this paper is to derive augmented Eilenberger equations including all the force terms 
that contribute to the charging in type-II superconductors 
and to clarify the charging mechanism of a vortex in an $s$-wave superconductor with a spherical Fermi surface.

It was previously pointed out that three forces contribute to the charging in superconductors: 
(i) the Lorentz force acting on the supercurrent \cite{London}, 
(ii) the force caused by the pair-potential gradient (PPG) \cite{Kopnin94,KopninText}, 
and (iii) the pressure arising from the slope in the density of states (DOS) \cite{Khomskii,Feigel'man}.

The existence of the Lorentz force acting on the supercurrent was first pointed out by London \cite{London,Kita09}.
He included a term corresponding to the Lorentz force acting on the supercurrent in his phenomenological equation.
Using this equation, it is found that the charging in superconductors due to the Hall effect occurs whenever a supercurrent flows. 
Subsequently, some authors have attempted to microscopically derive the transport equations for superconductors 
with the Lorentz force \cite{Betbeder,Aronov,Larkin,Kopnin94,Houghton}, 
which was microscopically recovered later in augmented quasiclassical equations of superconductivity in the Keldysh formalism 
as the next-to-leading-order contribution in the expansion of the Gor'kov equations \cite{Gor'kov}
in terms of the quasiclassical parameter 
$\delta \equiv 1/ k_{\rm F} \xi_0$ ($k_{\rm F}$: Fermi wavenumber, $\xi_0$: coherence length) \cite{Kita01}.
Using the augmented equations, the Hall coefficient of equilibrium supercurrent was derived \cite{Kita09}. 
Very recently, augmented Eilenberger equations incorporating the Lorentz force in the Matsubara formalism were derived \cite{Ueki} 
and used to calculate the vortex-core charging due to the Lorentz force 
as functions of the temperature \cite{Ueki} and magnetic field \cite{Kohno16,Kohno17}. 
More precisely, the component of the magnetic Lorentz force that balances the Hall electric field 
may be missing from the standard Eilenberger equations. 
It is known that the component of the magnetic Lorentz force that balances the hydrodynamic force 
exists in the Ginzburg--Landau (GL) equations \cite{Kato16}.

The PPG force was first discussed by Kopnin \cite{Kopnin94}. 
He derived a transport equation similar to the Boltzmann equation for clean type-II superconductors 
incorporating the Lorentz and PPG force terms to study the flux-flow Hall effect. 
In recent years, Arahata and Kato first 
included the Lorentz and PPG force terms in their augmented quasiclassical equations \cite{Arahata} 
as an extension of the standard quasiclassical equations of superconductivity in the Keldysh formalism \cite{Eliashberg}, 
and calculated flux-flow Hall conductivity numerically for $s$-wave superconductors with a cylindrical Fermi surface 
\cite{Arahata}. 
Subsequently, the charging mechanism was studied in an isolated vortex of an $s$-wave superconductor 
with a cylindrical Fermi surface 
based on augmented Eilenberger equations incorporating the Lorentz and PPG forces in the Matsubara formalism \cite{Ohuchi}. 
It was found that the PPG force contributes dominantly to charging in the core region of an isolated
vortex over a wide parameter range.
Very recently, the core charge in an isolated vortex of a chiral $p$-wave superconductor was calculated 
using the equations in the Matsubara formalism \cite{Masaki17}.

The charging mechanism of a superconducting vortex due to the pressure or chemical potential difference 
between the normal and superconducting states was first proposed by Khomskii and Freimuth \cite{Khomskii}. 
They regarded the core as a normal region 
and considered its chemical potential difference from the surroundings due to the particle--hole asymmetry 
in the DOS. 
For the general case of the DOS, 
Khomskii and Kusmartsev have also given a formula for the chemical potential difference between the normal and superconducting states due to the slope in the DOS
\cite{Khomskii92}. 
However, despite this, 
quasiclassical equations considering this pressure dependence have not yet been derived microscopically. 
More realistically, the pair potential in the vortex state of type-II superconductors 
is not in the form of a step function, as used in Ref. \citen{Khomskii}, 
but increases from the vortex center toward the outside 
and approaches the value of the pair potential in a homogeneous system. 
Moreover, since the chemical potential in the equilibrium vortex state is spatially homogeneous, 
it is not self-evident how this pressure term is added. 
Therefore, quasiclassical equations for superconductors still have room for improvement.

Here we derive the augmented Eilenberger equations in the Matsubara formalism with the Lorentz force, the PPG force, 
and the pressure due to the slope in the DOS by incorporating the first order of the quasiclassical parameter $\delta$. 
This pressure is called ``the SDOS pressure". 
Using it, we calculate the core charge in an isolated vortex of an $s$-wave superconductor with a spherical Fermi surface 
and study which of the three forces dominantly contributes to the charging.

This paper is organized as follows. 
In Sect. \ref{sec:II}, we derive augmented Eilenberger equations in the Matsubara formalism 
incorporating the Lorentz force, PPG force, and SDOS pressure.
In Sect. \ref{sec:III}, we present numerical results for the superconducting chemical potential 
and DOS in a homogeneous system and the core charge in an isolated vortex system.
In Sect. \ref{sec:IV}, we provide a brief summary.

\section{Augmented Eilenberger Equations \label{sec:II}}

\subsection{Matsubara Green's functions and Gor'kov equations \label{sec:A}}

We consider conduction electrons in static electromagnetic fields described by the Hamiltonian  
\begin{align}
&\hat{\cal H} = \int d \xi_1 \hat{\psi}^\dagger (\xi_1) \hat{\cal K}_1 \hat{\psi} (\xi_1) \notag \\ 
& \ \ \ + \frac{1}{2} \int d \xi_1 \int d \xi_2 {\cal V} ({\bm r}_1 -{\bm r}_2) 
\hat{\psi}^\dagger (\xi_1) \hat{\psi}^\dagger (\xi_2) \hat{\psi} (\xi_2) \hat{\psi} (\xi_1), 
\end{align}
where the variable $\xi$ is defined by $\xi \equiv ({\bm r}, \alpha)$ 
with ${\bm r}$ and $\alpha$ denoting the space and spin coordinates, respectively, 
$\hat{\psi}^\dagger (\xi)$ and $\hat{\psi} (\xi)$ are the creation and annihilation operators of the fermion field, respectively, 
$^\dagger$ denotes the Hermitian conjugate, 
and ${\cal V} ({\bm r}_1 - {\bm r}_2)$ is the interaction potential. 
Operator $\hat{\mathcal{K}}_1$ is defined by
\begin{equation}
\hat{\mathcal{K}}_1 \equiv \frac{1}{2 m} \left[ - i \hbar \frac{\partial}{\partial {\bm r}_1} - e {\bm A} ({\bm r}_1) \right]^2 + e \Phi ({\bm r}_1) - \mu,
\end{equation}
where $m$ is the electron mass, $e < 0$ is the electron charge, and $\mu$ is the chemical potential. 
$\Phi({\bm r})$ and ${\bm A}({\bm r})$ are the static scalar potential and vector potential, respectively, 
and static electromagnetic fields are expressed here in terms of them 
as ${\bm E}({\bm r})=-{\bm\nabla}\Phi({\bm r})$ and ${\bm B}({\bm r})={\bm\nabla}\times{\bm A}({\bm r})$. 
Next, we introduce the Heisenberg representations of the field operators in the Matsubara formalism as 
\begin{equation}
\left\{ \begin{array}{ll}
\displaystyle \hat\psi_1(1)\equiv e^{{\tau_1}\hat{\cal H}}\hat\psi (\xi_1)e^{-{\tau_1}\hat{\cal H}} \\
\displaystyle \hat\psi_2(1)\equiv e^{{\tau_1}\hat{\cal H}}\hat\psi^\dagger(\xi_1)e^{-{\tau_1}\hat{\cal H}}
\end{array}\right. , 
\end{equation}
where the argument $1$ in the round brackets denotes $1 \equiv (\xi_1,\tau_1)$, 
and the variable $\tau_1$ lies in the range $0 \le \tau_1 \le 1 / k_{\rm B} T$ 
with $k_{\rm B}$ and $T$ denoting the Boltzmann constant and temperature, respectively. 
Using them, we introduce the Matsubara Green's function
\begin{equation}
G_{ij} (1,2) \equiv - \langle T_\tau \hat{\psi}_i (1) \hat{\psi}_{3-j} (2) \rangle,
\end{equation}
where $T_\tau$ is the ``time''-ordering operator and $\langle \cdots \rangle$ denotes the grand-canonical average \cite{AGD,KitaText}. 
The elements satisfy \cite{KitaText}
\begin{equation}
G_{ij} (1,2) = - G_{3-j,3-i} (2, 1) = G_{ji}^* (\xi_2 \tau_1,\xi_1 \tau_2), \label{SRGij}
\end{equation}
where the superscript $^*$ denotes the complex conjugate.
The Matsubara Green's function can be expanded as \cite{KitaText}
\begin{equation}
G_{ij} (1,2) = k_{\rm B} T \sum_{n=-\infty}^\infty G_{ij} (\xi_1,\xi_2; \varepsilon_n) e^{- i \varepsilon_n (\tau_1-\tau_2)}, 
\label{FTGij}
\end{equation}
where $\varepsilon_n = (2n + 1) \pi k_{\rm B} T$ is the fermion Matsubara energy $(n=0, \pm 1, \ldots )$.
Separating the spin variable $\alpha = \uparrow, \downarrow$ from $\xi = ({\bm r}, \alpha)$, 
we introduce a new notation for each $G_{ij}$ 
as 
\begin{subequations}
\begin{align}
G_{11} (\xi_1,\xi_2; \varepsilon_n) &= G_{\alpha_1, \alpha_2} ({\bm r}_1, {\bm r}_2; \varepsilon_n), \\
G_{12} (\xi_1,\xi_2; \varepsilon_n) &= F_{\alpha_1, \alpha_2} ({\bm r}_1, {\bm r}_2; \varepsilon_n), \\
G_{21} (\xi_1,\xi_2; \varepsilon_n) &= - \bar{F}_{\alpha_1, \alpha_2} ({\bm r}_1, {\bm r}_2; \varepsilon_n), \\
G_{22} (\xi_1,\xi_2; \varepsilon_n) &= - \bar{G}_{\alpha_1, \alpha_2} ({\bm r}_1, {\bm r}_2; \varepsilon_n). 
\end{align} 
\end{subequations}
Subsequently, we express the spin degrees of freedom as the $2 \times 2$ matrix
\begin{align}
\underline{G} ({\bm r}_1, {\bm r}_2 ; \varepsilon_n) &\equiv
\begin{bmatrix}
G_{\uparrow \uparrow} ({\bm r}_1, {\bm r}_2 ; \varepsilon_n) & G_{\uparrow \downarrow} ({\bm r}_1, {\bm r}_2 ; \varepsilon_n) \\
G_{\downarrow \uparrow} ({\bm r}_1, {\bm r}_2 ; \varepsilon_n) & G_{\downarrow \downarrow} ({\bm r}_1, {\bm r}_2 ; \varepsilon_n)
\end{bmatrix}.
\end{align}
Then, we obtain the symmetry relations for $\underline{G}$ and $\underline{F}$ from Eqs. (\ref{SRGij}) and (\ref{FTGij}) as
\begin{subequations}
\begin{align}
\underline{G} ({\bm r}_1, {\bm r}_2 ; \varepsilon_n) &= \underline{G}^\dagger ({\bm r}_2, {\bm r}_1 ; - \varepsilon_n) = \underline{\bar{G}}^{\rm T} ({\bm r}_2, {\bm r}_1 ; - \varepsilon_n), \\
\underline{F} ({\bm r}_1, {\bm r}_2 ; \varepsilon_n) &= - \underline{\bar{F}}^\dagger ({\bm r}_2, {\bm r}_1 ; - \varepsilon_n) = - \underline{F}^{\rm T} ({\bm r}_2, {\bm r}_1 ; - \varepsilon_n),
\end{align}
\end{subequations}
where $^{\rm T}$ denotes the transpose.
It follows from these symmetry relations that $\underline{\bar{G}} ({\bm r}_1, {\bm r}_2 ; \varepsilon_n) = \underline{G}^* ({\bm r}_1, {\bm r}_2 ; \varepsilon_n)$
and $\underline{\bar{F}} ({\bm r}_1, {\bm r}_2 ; \varepsilon_n) = \underline{F}^* ({\bm r}_1, {\bm r}_2 ; \varepsilon_n)$ hold.
Using $\underline{G}$ and $\underline{F}$, we define a $4 \times 4$ Nambu matrix by
\begin{align}
\hat{G} ({\bm r}_1, {\bm r}_2 ; \varepsilon_n) &\equiv
\begin{bmatrix}
\underline{G} ({\bm r}_1, {\bm r}_2 ; \varepsilon_n) & \underline{F} ({\bm r}_1, {\bm r}_2 ; \varepsilon_n) \\
- \underline{F}^* ({\bm r}_1, {\bm r}_2 ; \varepsilon_n) & - \underline{G}^* ({\bm r}_1, {\bm r}_2 ; \varepsilon_n)
\end{bmatrix}.
\label{NambuGreen'sFunctions}
\end{align}
In the mean-field approximation, they satisfy the Gor'kov equations \cite{Gor'kov,KitaText}
\begin{align}
&
\begin{bmatrix}
( i \varepsilon_n - \hat{\mathcal{K}}_1 ) \underline{\sigma}_0  & \underline{0} \\
\underline{0} & ( i \varepsilon_n + \hat{\mathcal{K}}_1^* ) \underline{\sigma}_0
\end{bmatrix}
\hat{G} ({\bm r}_1, {\bm r}_2 ; \varepsilon_n) \notag \\
&- \int d^3 r_3 \hat{\mathcal{U}}_{\rm BdG} ({\bm r}_1, {\bm r}_3 ) \hat{G} ({\bm r}_3, {\bm r}_2 ; \varepsilon_n) = \hat{\delta} ({\bm r}_1-{\bm r}_2), 
\label{Gor'kovEq}
\end{align}
where $\underline{\sigma}_0$ and  $\underline{0}$ denote the $2 \times 2$ unit and zero matrices, respectively.
Matrix $\hat{\mathcal{U}}_{\rm BdG} ({\bm r}_1, {\bm r}_3 )$ denotes 
\begin{align}
\hat{\mathcal{U}}_{\rm BdG} ({\bm r}_1, {\bm r}_2) &\equiv
\begin{bmatrix}
\underline{\mathcal{U}}_{\rm HF} ({\bm r}_1, {\bm r}_2) & \underline{\Delta} ({\bm r}_1, {\bm r}_2) \\
- \underline{\Delta}^* ({\bm r}_1, {\bm r}_2) & - \underline{\mathcal{U}}_{\rm HF}^* ({\bm r}_1, {\bm r}_2) 
\end{bmatrix},
\label{UBdG}
\end{align}
where matrices $\underline{\mathcal{U}}_{\rm HF} ({\bm r}_1, {\bm r}_2)$ and $\underline{\Delta} ({\bm r}_1, {\bm r}_2)$ 
are the Hartree--Fock and pair potentials, respectively, defined by
\begin{align}
&\underline{\mathcal{U}}_{\rm HF} ({\bm r}_1, {\bm r}_2) \equiv \delta ({\bm r}_1 - {\bm r}_2) \underline{\sigma}_0 
{\rm Tr} \int d^3 r_3 {\cal V} ({\bm r}_1 - {\bm r}_3) \notag \\
& \ \ \ \times k_{\rm B} T \sum_{n = -\infty}^\infty
\underline{G} ({\bm r}_3, {\bm r}_3; \varepsilon_n) {\rm e}^{-i \varepsilon_n 0_-} \notag \\
& \ \ \ - {\cal V} ({\bm r}_1 - {\bm r}_2) k_{\rm B} T \sum_{n = -\infty}^\infty
\underline{G} ({\bm r}_1, {\bm r}_2; \varepsilon_n) {\rm e}^{-i \varepsilon_n 0_-}, \\
&\underline{\Delta} ({\bm r}_1, {\bm r}_2) \equiv {\cal V} ({\bm r}_1 - {\bm r}_2) 
k_{\rm B} T \sum_{n = -\infty}^\infty \underline{F} ({\bm r}_1, {\bm r}_2; \varepsilon_n), \label{Del}
\end{align}
where $0_-$ denotes an extra infinitesimal negative constant. 
Finally, matrix $\hat{\delta}$ on the right-hand side of Eq.\ (\ref{Gor'kovEq}) is defined by 
\begin{align}
\hat{\delta} ({\bm r}_1- {\bm r}_2) \equiv
\begin{bmatrix}
\delta ({\bm r}_1- {\bm r}_2) \underline{\sigma}_0 & \underline{0} \\
\underline{0} & \delta ({\bm r}_1-{\bm r}_2) \underline{\sigma}_0
\end{bmatrix}. 
\end{align}
%

\subsection{Gor'kov equations in the Wigner representation}

It is known that the original Wigner transform \cite{Wigner} breaks 
the gauge invariance with respect to the center-of-mass coordinate
when applied to the Green's functions of charged systems. 
Therefore, we use the gauge-covariant Wigner transform for the Matsubara Green's functions in Refs. \citen{Ueki} and  \citen{KitaText}, 
defined by
\begin{subequations}
\label{GIWT}
\begin{align}
&\hat{G} (\varepsilon_n, {\bm p}, {\bm r}_{12}) \notag \\
&\equiv \int d^3 \bar{r}_{12} e^{- i {\bm p} \cdot \bar{{\bm r}}_{12} / \hbar} \hat{\Gamma} ({\bm r}_{12}, {\bm r}_1) \hat{G} ({\bm r}_1, {\bm r}_2 ; \varepsilon_n) \hat{\Gamma} ({\bm r}_2, {\bm r}_{12}) \notag \\
&\equiv
\begin{bmatrix}
\underline{G} (\varepsilon_n, {\bm p}, {\bm r}_{12}) & \underline{F} (\varepsilon_n, {\bm p}, {\bm r}_{12}) \\
- \underline{F}^* (\varepsilon_n, - {\bm p}, {\bm r}_{12}) & - \underline{G}^* (\varepsilon_n, - {\bm p}, {\bm r}_{12})
\end{bmatrix}, 
\label{GIWT1}
\end{align}
with ${\bm r}_{12} \equiv ({\bm r}_1 + {\bm r}_2) / 2$ and $\bar{\bm r}_{12} \equiv {\bm r}_1 - {\bm r}_2$, the inverse of which is given by
\begin{align}
&\hat{G} ({\bm r}_1, {\bm r}_2 ; \varepsilon_n) \notag \\
&= \hat{\Gamma} ( {\bm r}_1,{\bm r}_{12}) \int \frac{d^3 p}{(2 \pi \hbar)^3} e^{i {\bm p} \cdot \bar{{\bm r}}_{12} / \hbar} \hat{G} (\varepsilon_n, {\bm p}, {\bm r}_{12}) 
\hat{\Gamma} ({\bm r}_{12},{\bm r}_2). 
\label{GIWT2}
\end{align}
\end{subequations}
Matrix $\hat{\Gamma}$ is given by 
\begin{align}
\hat{\Gamma} ({\bm r}_1, {\bm r}_2) &\equiv
\begin{bmatrix}
\underline{\sigma}_0 e^{i I ({\bm r}_1, {\bm r}_2)} & \underline{0} \\
\underline{0} & \underline{\sigma}_0 e^{- i I ({\bm r}_1, {\bm r}_2)}
\end{bmatrix}. 
\end{align}
Function $I ({\bm r}_1, {\bm r}_2)$ is the line integral defined by 
\begin{equation}
I ({\bm r}_1, {\bm r}_2) \equiv \frac{e}{\hbar} \int_{{\bm r}_2}^{{\bm r}_1} {\bm A} ({\bm s}) \cdot d {\bm s}, 
\label{I}
\end{equation}
where ${\bm s}$ denotes a straight-line path from ${\bm r}_2$ to ${\bm r}_1$. 
Similarly, we transform the mean-field potential in Eq. (\ref{UBdG}) as 
\begin{subequations}
\begin{align}
&\hat{\cal U}_{\rm BdG} ({\bm p}, {\bm r}_{12}) \notag \\
&\equiv \int d^3 \bar{r}_{12} e^{- i {\bm p} \cdot \bar{{\bm r}}_{12} / \hbar} \hat{\Gamma} ({\bm r}_{12}, {\bm r}_1) 
\hat{\cal U}_{\rm BdG} ({\bm r}_1, {\bm r}_2) \hat{\Gamma} ({\bm r}_2, {\bm r}_{12}) \notag \\
&\equiv
\begin{bmatrix}
\underline{\cal U}_{\rm HF} ({\bm p}, {\bm r}_{12}) & \underline{\Delta} ({\bm p}, {\bm r}_{12}) \\
- \underline{\Delta}^* (- {\bm p}, {\bm r}_{12}) & - \underline{\cal U}_{\rm HF}^* (- {\bm p}, {\bm r}_{12})
\end{bmatrix}, 
\end{align}
whose inverse is
\begin{align}
&\hat{\cal U}_{\rm BdG} ({\bm r}_1, {\bm r}_2) \notag \\
&= \hat{\Gamma} ( {\bm r}_1,{\bm r}_{12}) \int \frac{d^3 p}{(2 \pi \hbar)^3} e^{i {\bm p} \cdot \bar{{\bm r}}_{12} / \hbar} 
\hat{\cal U}_{\rm BdG} ({\bm p}, {\bm r}_{12}) 
\hat{\Gamma} ({\bm r}_{12},{\bm r}_2). 
\label{UIWT2}
\end{align}
\end{subequations}
Note the symmetry relations 
$\underline{\cal U}_{\rm HF} ({\bm p}, {\bm r}_{12}) = \underline{\cal U}_{\rm HF}^\dagger ({\bm p}, {\bm r}_{12})$ 
and $\underline{\Delta} ({\bm p}, {\bm r}_{12}) = - \underline{\Delta}^{\rm T} (-{\bm p}, {\bm r}_{12})$.

We consider the next-to-leading-order contribution in the expansion in terms of the quasiclassical parameter. 
Then, the Gor'kov equations in the Wigner representation are given as 
(see Appendices \ref{AppA} and \ref{AppB} 
for the derivation of the kinetic-energy and self-energy terms in the Wigner representation, respectively)
\begin{align}
&\left\{ i \varepsilon_n \hat{1} 
- \left[ \xi_{\bm p} - i \frac{\hbar {\bm v}}{2} \cdot {\bm \partial} - \frac{\hbar^2 {\bm \partial}^2}{8m^*} 
- \frac{i \hbar}{2} e {\bm E} ({\bm r}) \cdot \frac{\partial}{\partial {\bm p}} \right] \hat{\tau}_3 \right\} 
\hat{G} (\varepsilon_n, {\bm p}, {\bm r}) \notag \\
& \ \ \ \ \ - \hat{\Delta} ({\bm p}, {\bm r}) \circ \hat{G} (\varepsilon_n, {\bm p}, {\bm r}) \notag \\
& \ \ \ \ \ + \frac{i \hbar}{8} e {\bm v} \cdot \left[ {\bm B} ({\bm r}) \times \frac{\partial}{\partial {\bm p}} \right] 
\left[ 3 \hat{G} (\varepsilon_n, {\bm p}, {\bm r}) 
+ \hat{\tau}_3 \hat{G} (\varepsilon_n, {\bm p}, {\bm r}) \hat{\tau}_3 \right] = \hat{1}, 
\label{Gor'kovEqWigner1}
\end{align}
where $\xi_{\bm p}$ is defined by $\xi_{\bm p} \equiv \varepsilon_{\bm p} + e \Phi ({\bm r}) - \mu$ 
with $\varepsilon_{\bm p}$ denoting the single-particle energy, 
$m^*$ is the effective mass defined by $m^* \equiv p / v$, 
$\hat{1}$ denotes the $4 \times 4$ unit matrix, 
$\hat{\tau}_3$ is defined by 
\begin{align}
\hat{\tau}_3 \equiv
\begin{bmatrix}
\underline{\sigma}_0 & \underline{0} \\
\underline{0} & - \underline{\sigma}_0
\end{bmatrix}, 
\end{align}
${\bm \partial}$ is given by 
\begin{equation}
{\bm \partial} \equiv \left\{ \begin{array}{ll}
\displaystyle \frac{\partial}{\partial {\bm r}} & {\rm :on \ } \underline{G} {\ \rm or \ } \underline{G}^* \\
\displaystyle \frac{\partial}{\partial {\bm r}} - i \cfrac{2 e}{\hbar} {\bm A} ({\bm r}) & {\rm :on \ } \underline{F} \\
\displaystyle \frac{\partial}{\partial {\bm r}} + i \cfrac{2 e}{\hbar} {\bm A} ({\bm r}) & {\rm :on \ } \underline{F}^*
\end{array}\right. , \label{partial12}
\end{equation}
and the operator $\circ$ is defined by 
\begin{align}
\hat{a} ({\bm p}, {\bm r}) \circ \hat{b} ({\bm p}, {\bm r}) 
\equiv \hat{a} ({\bm p}, {\bm r}) 
\exp \left[ \frac{i \hbar}{2} \left( \overleftarrow{\bm \partial} \cdot \overrightarrow{\bm \partial}_{\bm p} 
- \overleftarrow{\bm \partial}_{\bm p} \cdot \overrightarrow{\bm \partial} \right) \right] 
\hat{b} ({\bm p}, {\bm r}). 
\end{align}
We take the Hermitian conjugate of Eq. (\ref{Gor'kovEqWigner1}), 
use the symmetries $\hat{\cal U}_{\rm BdG}^\dagger ({\bm p}, {\bm r}) = \hat{\cal U}_{\rm BdG} ({\bm p}, {\bm r})$ and 
$\hat{G}^\dagger (\varepsilon_n, {\bm p}, {\bm r}) = \hat{G} (- \varepsilon_n, {\bm p}, {\bm r})$,
and replace $\varepsilon_n \to - \varepsilon_n$ to obtain 
\begin{align}
&\hat{G} (\varepsilon_n, {\bm p}, {\bm r}) 
\left\{ i \varepsilon_n \hat{1} 
- \hat{\tau}_3 \left[ \xi_{\bm p} + i \frac{\hbar {\bm v}}{2} \cdot {\bm \partial} - \frac{\hbar^2 {\bm \partial}^2}{8m^*} 
+ \frac{i \hbar}{2} e {\bm E} ({\bm r}) \cdot \frac{\partial}{\partial {\bm p}} \right] \right\} \notag \\
& \ \ \ \ \ - \hat{G} (\varepsilon_n, {\bm p}, {\bm r}) \circ \hat{\Delta} ({\bm p}, {\bm r}) \notag \\
& \ \ \ \ \ - \frac{i \hbar}{8} e {\bm v} \cdot \left[ {\bm B} ({\bm r}) \times \frac{\partial}{\partial {\bm p}} \right] 
\left[ 3 \hat{G} (\varepsilon_n, {\bm p}, {\bm r}) 
+ \hat{\tau}_3 \hat{G} (\varepsilon_n, {\bm p}, {\bm r}) \hat{\tau}_3 \right] = \hat{1}. 
\label{Gor'kovEqWigner2}
\end{align} \label{Gor'kovEqWigner}%
We next operate $\hat{\tau}_3$ on the left- and right-hand sides of Eq. (\ref{Gor'kovEqWigner2}), 
and the resulting equation is subtracted from Eq. (\ref{Gor'kovEqWigner1}) and added to Eq. (\ref{Gor'kovEqWigner1}).  
Then, we obtain the following two equations: 
\begin{subequations}
\begin{align}
&\left[ i \varepsilon_n \hat{\tau}_3 - \hat{\Delta} ({\bm p}, {\bm r}) \hat{\tau}_3, 
\hat{\tau}_3 \hat{G}(\varepsilon_n, {\bm p}, {\bm r}) \right]_\circ 
+ i \hbar {\bm v} \cdot {\bm \partial} \hat{\tau}_3 \hat{G}(\varepsilon_n, {\bm p}, {\bm r}) \notag \\
&+ i \hbar e {\bm E} \cdot \frac{\partial}{\partial {\bm p}} \hat{\tau}_3 \hat{G}(\varepsilon_n, {\bm p}, {\bm r}) 
+ \frac{i \hbar}{2} e {\bm v} \cdot \left( {\bm B} \times \frac{\partial}{\partial {\bm p}} \right)
\left\{ \hat{\tau}_3, \hat{\tau}_3 \hat{G}(\varepsilon_n, {\bm p}, {\bm r}) \right\} \notag \\
&= \hat{0}, \label{L-RGEq}
\end{align}
\begin{align}
&\frac{1}{2} \left\{ i \varepsilon_n \hat{\tau}_3 - \hat{\Delta} ({\bm p}, {\bm r}) \hat{\tau}_3, 
\hat{\tau}_3 \hat{G}(\varepsilon_n, {\bm p}, {\bm r}) \right\}_\circ 
- \xi_{\bm p} \hat{\tau}_3 \hat{G}(\varepsilon_n, {\bm p}, {\bm r}) - \hat{1} \notag \\
&+ \frac{\hbar^2 {\bm \partial}^2}{8 m^*} \hat{\tau}_3 \hat{G}(\varepsilon_n, {\bm p}, {\bm r})
+ \frac{i \hbar}{8} e {\bm v} \cdot \left( {\bm B} \times \frac{\partial}{\partial {\bm p}} \right)
\left[ \hat{\tau}_3, \hat{\tau}_3 \hat{G}(\varepsilon_n, {\bm p}, {\bm r}) \right] \notag \\
&= \hat{0}, \label{L+RGEq}
\end{align}
\label{L+-RGEq}%
\end{subequations}
with $[ \hat{a}, \hat{b} ] \equiv \hat{a} \hat{b} - \hat{b} \hat{a}$, 
$[ \hat{a}, \hat{b} ]_\circ \equiv \hat{a} \circ \hat{b} - \hat{b} \circ \hat{a}$, 
$\{ \hat{a}, \hat{b} \} \equiv \hat{a} \hat{b} + \hat{b} \hat{a}$, and 
$\{ \hat{a}, \hat{b} \}_\circ \equiv \hat{a} \circ \hat{b} + \hat{b} \circ \hat{a}$.
Now, in terms of Eq. (\ref{GIWT1}), we introduce the quasiclassical Green's function 
\begin{align}
\hat{g} (\varepsilon_n, {\bm p}_{\rm F}, {\bm r}) 
&\equiv {\rm P} \int_{-\infty}^\infty \frac{d \xi_{\bm p}}{\pi} i \hat{\tau}_3 \hat{G}(\varepsilon_n, {\bm p}, {\bm r}) \notag \\
&\equiv 
\begin{bmatrix}
\underline{g} (\varepsilon_n, {\bm p}_{\rm F}, {\bm r})  
& - i \underline{f} (\varepsilon_n, {\bm p}_{\rm F}, {\bm r}) \\
-i \underline{f}^* (\varepsilon_n, - {\bm p}_{\rm F}, {\bm r}) 
& - \underline{g}^* (\varepsilon_n, - {\bm p}_{\rm F}, {\bm r})
\end{bmatrix}, \label{gdef}
\end{align}
where P denotes the principal value. 
It follows that the upper elements $\underline{g}$ and $\underline{f}$ satisfy
$\underline{g} (\varepsilon_n, {\bm p}_{\rm F}, {\bm r}) 
= - \underline{g}^\dagger (- \varepsilon_n, {\bm p}_{\rm F}, {\bm r})$, 
$\underline{f} (\varepsilon_n, {\bm p}_{\rm F}, {\bm r}) 
= - \underline{f}^{\rm T} (- \varepsilon_n, - {\bm p}_{\rm F}, {\bm r})$.
To derive the equation for $\hat{g}$ from Eq. (\ref{L-RGEq}), we express 
${\bm \partial}_{\bm p} = {\bm \partial}_{{\bm p}_\parallel} + {\bm v} (\partial / \partial \xi)$
with ${\bm p}_\parallel$ denoting the component on the energy surface $\xi = \xi_{\bm p}$, 
set ${\bm p} = {\bm p}_{\rm F}$ except for the argument of $\hat{G}$, 
integrate Eq. (\ref{L-RGEq}) over $- \varepsilon_{\rm c} \le \xi_{\bm p} \le \varepsilon_{\rm c}$, 
and use ${\bm v} \times {\bm \partial}_{{\bm p}_\parallel} = {\bm v} \times {\bm \partial}_{\bm p}$ and 
\begin{align}
{\rm P} \int_{- \infty}^\infty d \xi_{\bm p} \frac{\partial}{\partial \xi_{\bm p}} \hat{G} (\varepsilon_n, {\bm p}, {\bm r}) = \hat{0}.
\end{align}
We also neglect terms with $e {\bm E} \cdot {\bm \partial}_{{\bm p}_\parallel}$ 
and take the limit $\varepsilon_{\rm c} \to \infty$. 
We thereby obtain
\begin{align}
&\left[ i \varepsilon_n \hat{\tau}_3 - \hat{\Delta} ({\bm p}_{\rm F}, {\bm r}) \hat{\tau}_3, 
\hat{g} (\varepsilon_n, {\bm p}_{\rm F}, {\bm r}) \right]_\circ 
+ i \hbar {\bm v}_{\rm F} \cdot {\bm \partial} \hat{g} (\varepsilon_n, {\bm p}_{\rm F}, {\bm r}) \notag \\
& \ \ \ \ \ + \frac{i \hbar}{2} e {\bm v}_{\rm F} \cdot \left( {\bm B} \times \frac{\partial}{\partial {\bm p}_{\rm F}} \right)
\left\{ \hat{\tau}_3, \hat{g} (\varepsilon_n, {\bm p}_{\rm F}, {\bm r}) \right\} = \hat{0}, \label{AQCEq}
\end{align}
where ${\bm v}_{\rm F}$ is the Fermi velocity.
The term of the PPG ${\bm \partial} \Delta$ in Eq. (\ref{AQCEq}) 
is called the PPG force in Ref. \citen{Ohuchi}, 
the term of the magnetic field ${\bm B}$ is the magnetic Lorentz force term \cite{Kita01}, 
and the other terms are the main part of the standard Eilenberger equations \cite{Eilenberger}. 
We also include the effects of impurity scattering in the self-consistent Born approximation by \cite{Ueki}
\begin{align}
\hat{\sigma}_{\rm imp} (\varepsilon_n, {\bm r}) 
\equiv - i \frac{\hbar}{2 \tau} \int \frac{d \Omega_{\bf p}}{4 \pi} \hat{g} (\varepsilon_n, {\bm p}_{\rm F}, {\bm r}) \hat{\tau}_3,
\end{align}
where $\Omega_{\bm p}$ denotes the solid angle of momentum.
Then, the augmented quasiclassical equations in the Matsubara formalism are given by 
\begin{align}
&\left[ i \varepsilon_n \hat{\tau}_3 - \hat{\Delta} ({\bm p}_{\rm F}, {\bm r}) \hat{\tau}_3 
- \hat{\sigma}_{\rm imp} (\varepsilon_n, {\bm r}) \hat{\tau}_3, 
\hat{g} (\varepsilon_n, {\bm p}_{\rm F}, {\bm r}) \right]_\circ \notag \\
& \ \ \ + i \hbar {\bm v}_{\rm F} \cdot {\bm \partial} \hat{g} (\varepsilon_n, {\bm p}_{\rm F}, {\bm r}) 
+ \frac{i \hbar}{2} e {\bm v}_{\rm F} \cdot \left( {\bm B} \times \frac{\partial}{\partial {\bm p}_{\rm F}} \right) 
\left\{ \hat{\tau}_3, \hat{g} (\varepsilon_n, {\bm p}_{\rm F}, {\bm r}) \right\} \notag \\
& \ \ \ = \hat{0}. \label{AEEq}
\end{align}

Applying the same procedure to Eq. (\ref{AQCEq}), we obtain the equation for 
\begin{align}
\hat{g}^{(1)} (\varepsilon_n, {\bm p}_{\rm F}, {\bm r}) 
\equiv {\rm P} \int_{-\infty}^\infty \frac{d \xi_{\bm p}}{\pi} 
i \left[ \xi_{\bm p} \hat{\tau}_3 \hat{G}(\varepsilon_n, {\bm p}, {\bm r}) + \hat{1} \right] \label{g(1)def}
\end{align}
as 
\begin{align}
&\hat{g}^{(1)} (\varepsilon_n, {\bm p}_{\rm F}, {\bm r}) 
= \frac{1}{2} \left\{ i \varepsilon_n \hat{\tau}_3 - \hat{\Delta} ({\bm p}_{\rm F}, {\bm r}) \hat{\tau}_3, 
\hat{g} (\varepsilon_n, {\bm p}_{\rm F}, {\bm r})  \right\}_\circ \notag \\
& \ \ \ \ \ + \frac{\hbar^2 {\bm \partial}^2}{8 m^*} \hat{g} (\varepsilon_n, {\bm p}_{\rm F}, {\bm r})  
+ \frac{i \hbar}{8} e {\bm v}_{\rm F} \cdot \left( {\bm B} \times \frac{\partial}{\partial {\bm p}_{\rm F}} \right) 
\left[ \hat{\tau}_3, \hat{g} (\varepsilon_n, {\bm p}_{\rm F}, {\bm r}) \right]. \label{g(1)4}
\end{align}
Neglecting the second and third terms in Eq. (\ref{g(1)4}) to take the leading order as 
\begin{align}
\hat{g}^{(1)} (\varepsilon_n, {\bm p}_{\rm F}, {\bm r}) 
\approx \frac{1}{2} \left\{ i \varepsilon_n \hat{\tau}_3 - \hat{\Delta} ({\bm p}_{\rm F}, {\bm r}) \hat{\tau}_3, 
\hat{g} (\varepsilon_n, {\bm p}_{\rm F}, {\bm r})  \right\}, \label{g(1)2}
\end{align}
we use it to calculate the terms of the slope in the DOS.

\subsection{Local density of states}

Let us introduce the local density of states (LDOS) as 
\begin{align}
&N_{\rm s} (\varepsilon, {\bm r}) \equiv - {\rm Tr} \int \frac{d^3 p}{(2 \pi \hbar)^3} \frac{1}{2 \pi} {\rm Im} 
\underline{G}^{\rm R} (\varepsilon, {\bm p}, {\bm r}) \notag \\
&= - {\rm Tr} \int_{- \infty}^\infty d \xi_{\bm p} N(\xi_{\bm p} + \mu -e \Phi ({\bm r})) 
\int \frac{d \Omega_{\bm p}}{4 \pi} \frac{1}{2 \pi} {\rm Im} 
\underline{G}^{\rm R} (\varepsilon, {\bm p}, {\bm r}), \label{Ns}
\end{align}
where $\underline{G}^{\rm R} (\varepsilon, {\bm p}, {\bm r}) 
\equiv \underline{G} (\varepsilon_n \to - i \varepsilon + \eta, {\bm p}, {\bm r})$ is the retarded Green's function
with $\eta$ denoting the infinitesimal positive constant, 
and $N (\epsilon)$ is the normal DOS defined by
\begin{align}
N (\epsilon) \equiv \int \frac{d^3 p}{(2 \pi \hbar)^3} \delta (\epsilon - \varepsilon_{\bm p}). 
\end{align}
We here assume that (i) the superconducting DOS approaches the normal one as the single-particle energy increases 
and (ii) the energy variation of the normal DOS is slow. 
In this case, 
the superconducting DOS in Eq. (\ref{Ns}) may be expressed in terms of $\underline{g}$ and $\underline{g}^{(1)}$. 
To see this, let us expand $N (\epsilon)$ at $\epsilon = \mu - e \Phi ({\bm r})$ as 
$N (\xi_{\bm p} + \mu - e \Phi ({\bm r})) \approx N (\mu - e \Phi ({\bm r})) + N' (\mu - e \Phi ({\bm r})) \xi_{\bm p}$. 
Using $N'(\mu_{\rm n}) \Delta_0 / N(\mu_{\rm n}) = O (\delta)$, $\delta \mu / \Delta_0 = O (\delta)$, 
and $|e| \Phi / \Delta_0 = O (\delta)$, 
we obtain $N (\mu - e \Phi ({\bm r})) = N(\mu_{\rm n}) [ 1 + O (\delta^2)]$.
Here $\Delta_0$ denotes the energy gap at zero temperature, 
and $\delta \mu$ is the chemical potential difference between the normal and superconducting states defined by 
$\delta \mu \equiv \mu-\mu_{\rm n}$ with $\mu_{\rm n}$ denoting the chemical potential in the normal state.
Thus, we rewrite the expansion for $N (\xi_{\bm p} + \mu - e \Phi ({\bm r}))$ as
\begin{align}
N (\xi_{\bm p} + \mu - e \Phi ({\bm r})) \approx N (\mu_{\rm n}) + N' (\mu_{\rm n}) \xi_{\bm p}. 
\label{Nexpand}
\end{align}
Substituting it into Eq. (\ref{Ns}) and using Eqs. (\ref{gdef}), (\ref{g(1)def}), and (\ref{g(1)2}),  
we obtain the superconducting LDOS as 
\begin{align}
&N_{\rm s} (\varepsilon, {\bm r}) \notag \\
&\approx
\frac{N(\mu_{\rm n})}{2} 
{\rm Tr} \int \cfrac{d \Omega_{\bm p}}{4 \pi} \Bigg\{ {\rm Re} \underline{g}^{\rm R} (\varepsilon, {\bm p}_{\rm F}, {\bm r}) 
+ \cfrac{N'(\mu_{\rm n})}{N (\mu_{\rm n})} 
\varepsilon {\rm Re} \underline{g}^{\rm R} (\varepsilon, {\bm p}_{\rm F}, {\bm r}) \notag \\
& \ \ \  
+ \frac{1}{2} \cfrac{N'(\mu_{\rm n})}{N (\mu_{\rm n})} {\rm Im} 
\left[ \underline{\Delta} ({\bm p}_{\rm F}, {\bm r}) \underline{\bar{f}}^{\rm R} (\varepsilon, {\bm p}_{\rm F}, {\bm r}) 
+ \underline{f}^{\rm R} (\varepsilon, {\bm p}_{\rm F}, {\bm r}) \underline{\bar{\Delta}} ({\bm p}_{\rm F}, {\bm r}) \right]  
\Bigg\} \notag \\
& \ \ \ \times \theta (| \tilde{\varepsilon}_{\rm c} | - | \varepsilon |) 
+ N (\varepsilon + \mu - e \Phi ({\bm r})) \theta ( | \varepsilon | - | \tilde{\varepsilon}_{\rm c} | ){\color{cyan},} \label{Ns3} 
\end{align}
where the retarded Green's functions and barred functions in the Keldysh formalism 
are defined generally by $\underline{g}^{\rm R} (\varepsilon, {\bm p}_{\rm F}, {\bm r}) 
\equiv \underline{g} (\varepsilon_n \to - i \varepsilon + \eta, {\bm p}_{\rm F}, {\bm r})$
and $\underline{\bar{g}}^{\rm R} (\varepsilon, {\bm p}_{\rm F}, {\bm r}) 
\equiv \underline{g}^{\rm R}{}^* (-\varepsilon, -{\bm p}_{\rm F}, {\bm r})$, respectively.
We should determine the cutoff energy 
$\tilde{\varepsilon}_{\rm c}>0$ to satisfy  
\begin{align}
&\int_{- \tilde{\varepsilon}_{\rm c}}^{\tilde{\varepsilon}_{\rm c}} N_{\rm s} (\varepsilon, {\bm r}) d \varepsilon
= \int_{- \tilde{\varepsilon}_{\rm c}}^{\tilde{\varepsilon}_{\rm c}} N (\varepsilon + \mu - e \Phi) d \varepsilon. 
\end{align}
%

\subsection{Pair potential \label{subsec:2.7}}

We rewrite the self-consistency equation for the pair potential using the quasiclassical Green's function.
Substituting Eqs. (\ref{GIWT2}), (\ref{UIWT2}), and 
\begin{equation}
\mathcal{V} (| \bar{\bm r}_{12} |) = \int \frac{d p^3}{(2 \pi \hbar)^3} 
\mathcal{V}_p {\rm e}^{i {\bm p} \cdot \bar{\bm r}_{12} / \hbar}, 
\end{equation}
into Eq. (\ref{Del}), 
we then obtain $\underline{\Delta} ({\bm p}, {\bm r}_{12})$ as 
\begin{equation}
\underline{\Delta} ({\bm p}, {\bm r}_{12}) = \int \frac{d {p'}^3}{(2 \pi \hbar)^3} 
\mathcal{V}_{|{\bm p} - {\bm p}'|} k_{\rm B} T \sum_{n = - \infty}^\infty 
\underline{F} (\varepsilon_n, {\bm p}', {\bm r}_{12}). \label{Deltapr}
\end{equation}
Expanding the interaction $\mathcal{V}_{| {\bm p} - {\bm p}' |}$ 
with respect to the surface harmonics $Y_{lm} (\hat{\bm p})$
as
\begin{equation}
\mathcal{V}_{| {\bm p} - {\bm p}' |} = \sum_{l = 0}^\infty \mathcal{V}_l ({\bm p}, {\bm p}') 
\sum_{m=-l}^l 4 \pi Y_{lm} (\hat{\bm p}) Y_{lm}^* (\hat{\bm p}'), 
\end{equation}
we also assume that a single $l$ is relevant. 
Equation (\ref{Deltapr}) then becomes
\begin{align}
&\underline{\Delta} ({\bm p}, {\bm r}) 
= \int_{- \infty}^\infty d \xi_{\bm p'} N(\xi_{\bm p'} + \mu - e \Phi ({\bm r})) \int \frac{d \Omega_{{\bm p}'}}{4 \pi} \notag \\
&\times \mathcal{V}_l ({\bm p}, {\bm p}') \sum_{m=-l}^l 4\pi Y_{lm} (\hat{\bm p}) Y_{lm}^* (\hat{\bm p}') 
k_{\rm B} T \sum_{n=-\infty}^\infty \underline{F} (\varepsilon_n, {\bm p}', {\bm r}). \label{Deltapr2}
\end{align}
Assuming the constant and weak-coupling interaction, 
we can rewrite the interaction potential as $\mathcal{V}_l ({\bm p}, {\bm p}') 
= \mathcal{V}_l^{\rm (eff)} \theta (\varepsilon_{\rm c} -|\xi_{\bm p}|) \theta (\varepsilon_{\rm c} -|\xi_{{\bm p}'}|)$
with the constant potential $\mathcal{V}_l^{\rm (eff)}$ and cutoff energy $\varepsilon_{\rm c}$ \cite{KitaText}. 
We can also rewrite Eq. (\ref{Deltapr2}) as 
\begin{align}
&\underline{\Delta} ({\bm p}, {\bm r}) 
= \int_{- \varepsilon_{\rm c}}^{\varepsilon_{\rm c}} d \xi_{\bm p'} N(\xi_{\bm p'} + \mu - e \Phi ({\bm r})) \int \frac{d \Omega_{{\bm p}'}}{4 \pi} \notag \\
& \ \ \ \times \mathcal{V}_l^{\rm (eff)} \sum_{m=-l}^l 4\pi Y_{lm} (\hat{\bm p}) Y_{lm}^* (\hat{\bm p}') 
k_{\rm B} T \sum_{n=- n_{\rm c}}^{n_{\rm c}} \underline{F} (\varepsilon_n, {\bm p}', {\bm r}), 
\end{align}
where the cutoff $n_{\rm c}$ is determined from $(2 n_{\rm c} + 1) \pi k_{\rm B} T = \varepsilon_{\rm c}$ \cite{KitaText}. 
Using Eqs. (\ref{gdef}), (\ref{g(1)def}), (\ref{g(1)2}), and (\ref{Nexpand}), 
we see that only the value at ${\bm p} = {\bm p}_{\rm F}$ 
contributes to $\underline{\Delta} ({\bm p}, {\bm r})$ as
\begin{align}
&\underline{\Delta} ({\bm p}_{\rm F}, {\bm r}) 
\approx - \mathcal{V}_l^{\rm (eff)} N (\mu_{\rm n}) 
\int \frac{d \Omega_{{\bm p}'}}{4 \pi}  
\sum_{m=-l}^l 4\pi Y_{lm} (\hat{\bm p}) Y_{lm}^* (\hat{\bm p}') \notag \\
& \ \ \ \times \pi k_{\rm B} T \sum_{n = - n_{\rm c}}^{n_{\rm c}} 
\Bigg\{ \underline{f} (\varepsilon_n, {\bm p}_{\rm F}', {\bm r}) 
- \frac{i}{2} \frac{N'(\mu_{\rm n})}{N(\mu_{\rm n})}
\Big[ \underline{\Delta} ({\bm p}_{\rm F}', {\bm r}) \underline{\bar{g}} (\varepsilon_n, {\bm p}_{\rm F}', {\bm r}) \notag \\
& \ \ \ - \underline{g} (\varepsilon_n, {\bm p}_{\rm F}', {\bm r}) \underline{\Delta} ({\bm p}_{\rm F}', {\bm r}) \Big] 
\Bigg\}{\color{cyan},} 
\end{align}
where the barred functions in the Matsubara formalism are defined generally by 
$\underline{\bar{g}} (\varepsilon_n, {\bm p}_{\rm F}, {\bm r}) \equiv \underline{g}^* (\varepsilon_n, -{\bm p}_{\rm F}, {\bm r})$.
Expanding $\underline{\Delta} ({\bf p}_{\rm F}, {\bf r})$ with respect to the surface harmonics as
\begin{equation}
\underline{\Delta} ({\bm p}_{\rm F}, {\bm r}) 
= \sum_{m=-l}^l \underline{\Delta}_{lm} ({\bf r}) \sqrt{4 \pi} Y_{lm} (\hat{\bm p}), 
\end{equation}
the self-consistency equation for the pair potential is given by
\begin{align}
&\underline{\Delta}_{lm} ({\bf r})
= 2 \pi {\rm g}_0 k_{\rm B} T \sum_{n=0}^{n_{\rm c}} \int \frac{d \Omega_{\bf p}}{4 \pi} \sqrt{4 \pi} Y_{lm}^* (\hat{\bm p}) 
\Bigg\{ \underline{f} (\varepsilon_n, {\bm p}_{\rm F}, {\bm r})
\notag \\
& \ \ \ - \frac{i}{2} \frac{N'(\mu_{\rm n})}{N(\mu_{\rm n})}
\Big[ \underline{\Delta} ({\bm p}_{\rm F}, {\bm r}) \underline{\bar{g}} (\varepsilon_n, {\bm p}_{\rm F}, {\bm r}) 
- \underline{g} (\varepsilon_n, {\bm p}_{\rm F}, {\bm r}) \underline{\Delta} ({\bm p}_{\rm F}, {\bm r}) \Big] 
\Bigg\}, \label{GapEq}
\end{align}
where ${\rm g}_0 \equiv - \mathcal{V}_l^{\rm (eff)} N (\mu_{\rm n})$ denotes the coupling constant. 
Neglecting the spin magnetism as $\underline{g} = g \underline{\sigma}_0$,
the gap equation [Eq. (\ref{GapEq})] becomes the same as that in the standard Eilenberger equations.

\subsection{Charge and current densities}
We here express the charge density using the quasiclassical Green's function. 
First, we introduce the electron density $n ({\bm r})$ as 
\begin{align}
n ({\bm r}) &= k_{\rm B} T {\rm Tr} \sum_{n = - \infty}^\infty 
\underline{G} ({\bm r}, {\bm r}; \varepsilon_n) {\rm e}^{- i \varepsilon_n 0_-} 
= 2 \int_{- \infty}^\infty d \varepsilon \frac{N_{\rm s} (\varepsilon, {\bm r})}{{\rm e}^{\varepsilon / k_{\rm B} T} + 1}. \label{n}
\end{align}
Substituting Eq. (\ref{Ns3}) into Eq. (\ref{n}), the electron density is expressible in terms of 
$\underline{g}^{\rm R}$ and $\underline{g}^{{\rm R}(1)}$ as 
\begin{align}
&n ({\bm r}) 
\approx N (\mu_{\rm n}) {\rm Tr} \int_{- \tilde{\varepsilon}_{\rm c}}^{\tilde{\varepsilon}_{\rm c}} 
d \varepsilon \frac{1}{{\rm e}^{\varepsilon / k_{\rm B} T} + 1} \notag \\
& \ \ \times \int \frac{d \Omega_{\bm p}}{4 \pi} 
\left[ {\rm Re} \underline{g}^{\rm R} (\varepsilon, {\bm p}_{\rm F}, {\bm r}) 
+ \frac{N' (\mu_{\rm n})}{N (\mu_{\rm n})} {\rm Re} \underline{g}^{{\rm R}(1)} (\varepsilon, {\bm p}_{\rm F}, {\bm r}) \right] 
\notag \\
& \ \ + 2 \int_{- \infty}^\infty d \varepsilon 
\frac{N(\varepsilon + \mu - e \Phi ({\bm r}))}{{\rm e}^{\varepsilon / k_{\rm B} T} + 1} 
- 2 \int_{- \tilde{\varepsilon}_{\rm c}}^{\tilde{\varepsilon}_{\rm c}} d \varepsilon 
\frac{N(\varepsilon + \mu - e \Phi ({\bm r}))}{{\rm e}^{\varepsilon / k_{\rm B} T} + 1}, 
\end{align}
We also introduce the electron density in the normal state $n_{\rm n}$ as
\begin{align}
n_{\rm n} = 2 \int_{- \infty}^\infty d \varepsilon \frac{N(\varepsilon + \mu_{\rm n})}{{\rm e}^{\varepsilon / k_{\rm B} T} + 1}. 
\end{align}
Using it, the charge density $\rho ({\bm r}) = e n ({\bm r}) - e n_{\rm n}$ is given as 
\begin{align}
&\rho ({\bm r}) \approx e N(\mu_{\rm n}) {\rm Tr} \int_{-\tilde{\varepsilon}_{\rm c}}^{\tilde{\varepsilon}_{\rm c}} 
d \varepsilon \frac{1}{{\rm e}^{\varepsilon / k_{\rm B} T} + 1} \notag \\
&\times \int \frac{d \Omega_{\bm p}}{4 \pi}
\left[ {\rm Re} \underline{g}^{\rm R} (\varepsilon, {\bm p}_{\rm F}, {\bm r}) 
+ \frac{N' (\mu_{\rm n})}{N (\mu_{\rm n})} {\rm Re} \underline{g}^{{\rm R}(1)} (\varepsilon, {\bm p}_{\rm F}, {\bm r}) \right] \notag \\
&- 2 e \int_{- \tilde{\varepsilon}_{\rm c}}^{\tilde{\varepsilon}_{\rm c}} d \varepsilon 
\frac{N(\varepsilon + \mu - e \Phi ({\bm r}))}{{\rm e}^{\varepsilon / k_{\rm B} T} + 1} \notag \\
&+ 2 e \int_{- \infty}^\infty d \varepsilon N (\varepsilon) 
\left[ \frac{1}{{\rm e}^{(\varepsilon + e \Phi ({\bm r}) - \delta \mu - \mu_{\rm n}) / k_{\rm B} T}+1} 
- \frac{1}{{\rm e}^{(\varepsilon-\mu_{\rm n}) / k_{\rm B} T}+1} \right]. \label{rho} 
\end{align}
Let us carry out a perturbation expansion with respect to the Lorentz and PPG forces as 
$\underline{g}^{\rm R} = \underline{g}_0^{\rm R} + \underline{g}_1^{\rm R} \cdots$ and 
$\underline{g}^{{\rm R}(1)} = \underline{g}_0^{{\rm R}(1)} + \underline{g}_1^{{\rm R}(1)} \cdots$ \cite{Kita09,Ueki,Ohuchi}, 
which is performed up to the first order in the quasiclassical parameter $\delta$ below. 
We also use Eq. (\ref{g(1)2}), the $\underline{g}_0^{\rm R}$ and gap equation [Eq. (\ref{GapEq})] in the standard Eilenberger equations, 
and the following approximation for the distribution function: 
\begin{align}
&\frac{1}{{\rm e}^{(\varepsilon + e \Phi ({\bm r}) - \delta \mu - \mu_{\rm n}) / k_{\rm B} T}+1} \notag \\
& \ \ \ \approx \frac{1}{{\rm e}^{(\varepsilon-\mu_{\rm n}) / k_{\rm B} T}+1} 
+ \frac{d}{d \varepsilon} \frac{1}{{\rm e}^{(\varepsilon-\mu_{\rm n}) / k_{\rm B} T}+1} [e \Phi ({\bm r}) - \delta \mu ]. 
\end{align}
Then, we obtain the formula for the charge density as 
\begin{align}
&\rho ({\bm r}) \approx 
2 \pi k_{\rm B} T e N(\mu_{\rm n}) {\rm Tr} \sum_{n=0}^{\tilde{n}_{\rm c}} \int \frac{d \Omega_{\bm p}}{4 \pi} 
{\rm Im} \underline{g}_1 (\varepsilon_n, {\bm p}_{\rm F}, {\bm r}) \notag \\
& \ \ \ + e \frac{N' (\mu_{\rm n})}{N (\mu_{\rm n})} \int_{-\tilde{\varepsilon}_{\rm c}}^{\tilde{\varepsilon}_{\rm c}} d \varepsilon 
\frac{\varepsilon}{{\rm e}^{\varepsilon / k_{\rm B} T} + 1} 
\left[ N_{{\rm s}0} (\varepsilon, {\bm r}) - 2 N (\mu_{\rm n}) \right] \notag \\
& \ \ \ - (-1)^l c e N(\mu_{\rm n}) \frac{N' (\mu_{\rm n})}{N (\mu_{\rm n})} 
{\rm Tr} \sum_{m=-l}^l | \underline{\Delta}_{lm} ({\bm r}) |^2 \notag \\
& \ \ \ - 2 e N (\mu_{\rm n}) [ e \Phi ({\bm r}) - \delta \mu ], \label{rho2}
\end{align}
where the cutoff $\tilde{n}_{\rm c}$ is obtained from $(2 \tilde{n}_{\rm c} + 1) \pi k_{\rm B} T = \tilde{\varepsilon}_{\rm c}$, 
the coefficient $c$ is defined by 
\begin{align}
c \equiv \int_{-\tilde{\varepsilon}_{\rm c}}^{\tilde{\varepsilon}_{\rm c}} d \varepsilon \frac{1}{2\varepsilon} 
\tanh \frac{\varepsilon}{2 k_{\rm B} T_{\rm c}}, 
\end{align}
with $T_{\rm c}$ denoting the superconducting transition temperature at zero magnetic field, 
and $N_{{\rm s}0} (\varepsilon, {\bm r})$ is the LDOS obtained from the standard Eilenberger equations defined by
\begin{align}
N_{{\rm s}0} (\varepsilon, {\bm r}) \equiv \frac{N(\mu_{\rm n})}{2} {\rm Tr} \int \frac{d \Omega_{\bm p}}{4 \pi} 
{\rm Re} \underline{g}_0^{\rm R} (\varepsilon, {\bm p}_{\rm F}, {\bm r}). 
\end{align}
Using Gauss's law ${\bm \nabla} \cdot {\bm E} = \rho / \epsilon_0$, 
we obtain an equation for the electric field as
\begin{align}
&- \lambda_{\rm TF}^2 {\bm \nabla}^2 {\bm E} ({\bm r}) + {\bm E} ({\bm r}) \notag \\
& \ \ \ = - \frac{\pi k_{\rm B} T}{e} {\bm \nabla} {\rm Tr} \sum_{n=0}^{\tilde{n}_{\rm c}} \int \frac{d \Omega_{\bm p}}{4 \pi} 
{\rm Im} \underline{g}_1 (\varepsilon_n, {\bm p}_{\rm F}, {\bm r}) \notag \\
& \ \ \ -\frac{1}{e} \frac{N'(\mu_{\rm n})}{N(\mu_{\rm n})} 
\int_{-\tilde{\varepsilon}_{\rm c}}^{\tilde{\varepsilon}_{\rm c}} d \varepsilon 
\frac{\varepsilon}{{\rm e}^{\varepsilon / k_{\rm B} T} + 1} 
{\bm \nabla} \frac{N_{{\rm s}0} (\varepsilon, {\bm r})}{N(\mu_{\rm n})} \notag \\
& \ \ \ + (-1)^l \frac{c}{2e} \frac{N'(\mu_{\rm n})}{N(\mu_{\rm n})} {\bm \nabla} {\rm Tr} 
\sum_{m=-l}^l |\underline{\Delta}_{lm} ({\bm r})|^2, \label{EEq}
\end{align}
where $\lambda_{\rm TF} \equiv \sqrt{\epsilon_0 / 2 e^2 N (\mu_{\rm n})}$ is the Thomas--Fermi screening length. 
This expression includes the same screening effect as that in Refs. \citen{Eliashberg,Serene}, and \citen{Eschrig}.

\begin{figure}[t]
        \begin{center}
                \includegraphics[width=0.9\linewidth]{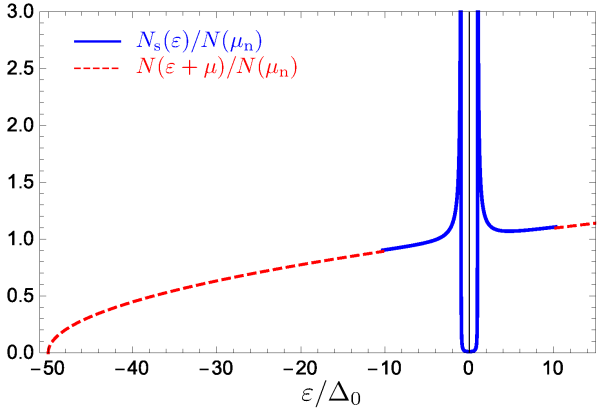}
                \end{center}
\caption{(Color online) Superconducting DOS $N_{\rm s} (\varepsilon)$ (blue solid line) 
and normal DOS $N (\varepsilon)$ (red dashed line) 
in units of $N(\mu_{\rm n})$ over $-50 \Delta_0 \le \varepsilon \le 10 \Delta_0$ at $T = 0.1T_{\rm c}$.}
\label{fig1}
\end{figure}
\begin{figure}[t]
        \begin{center}
                \includegraphics[width=0.9\linewidth]{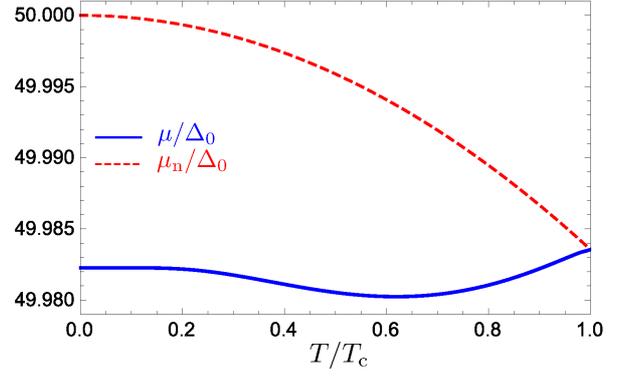}
                \end{center}
\caption{(Color online) Superconducting chemical potential $\mu$ (blue solid line) 
and normal chemical potential $\mu_{\rm n}$ (red dashed line) in units of $\Delta_0$
as a function of temperature.}
\label{fig2}
\end{figure}

On the other hand, using $\langle {\bm v}_{\rm F} \rangle_{\rm F} = {\bm 0}$, 
we obtain the formula for the current density by the same procedure as 
\begin{align}
{\bm j} ({\bm r}) 
&\approx e N (\mu_{\rm n}) {\rm Tr} \int_{-\tilde{\varepsilon}_{\rm c}}^{\tilde{\varepsilon}_{\rm c}} d \varepsilon 
\frac{1}{{\rm e}^{\varepsilon / k_{\rm B} T} + 1} \int \frac{d \Omega_{\bm p}}{4 \pi} {\bm v}_{\rm F} 
\Bigg[ {\rm Re} \underline{g}_0^{\rm R} (\varepsilon, {\bm p}_{\rm F}, {\bm r}) \notag \\
& \ \ \ + {\rm Re} \underline{g}_1^{\rm R} (\varepsilon, {\bm p}_{\rm F}, {\bm r}) 
+ \frac{N'(\mu_{\rm n})}{N(\mu_{\rm n})} {\rm Re} \underline{g}_0^{{\rm R}(1)} (\varepsilon, {\bm p}_{\rm F}, {\bm r}) 
\Bigg]. \label{j}
\end{align}
The second and third terms in Eq. (\ref{j}) with respect to $\underline{g}_1^{\rm R}$ and $\underline{g}_0^{{\rm R}(1)}$ 
are the correction terms due to the spatial variation of the electron density. 
Hence, neglecting the second and third terms in the above expression to take the leading order, 
we use the formula for the current density as
\begin{align}
{\bm j} ({\bm r}) &\approx e N (\mu_{\rm n}) 
{\rm Tr} \int_{-\tilde{\varepsilon}_{\rm c}}^{\tilde{\varepsilon}_{\rm c}} d \varepsilon 
\frac{1}{{\rm e}^{\varepsilon / k_{\rm B} T} + 1} \int \frac{d \Omega_{\bm p}}{4 \pi} {\bm v}_{\rm F} 
{\rm Re} \underline{g}_0^{\rm R} (\varepsilon, {\bm p}_{\rm F}, {\bm r}) \notag \\
&= 2 \pi k_{\rm B} T e N (\mu_{\rm n}) {\rm Tr} \sum_{n=0}^{\tilde{n}_{\rm c}} 
\int \frac{d \Omega_{\bm p}}{4 \pi} {\bm v}_{\rm F} 
{\rm Im} \underline{g}_0 (\varepsilon_n, {\bm p}_{\rm F}, {\bm r}). 
\end{align}
%

\subsection{Chemical potential}
We also obtain the expression for the chemical potential $\mu$ from Eq. (\ref{rho2}) as 
\begin{align}
&\mu = \mu_{\rm n} - \pi k_{\rm B} T {\rm Tr} \sum_{n=0}^{\tilde{n}_{\rm c}} \int \frac{d \Omega_{\bm p}}{4 \pi} 
\frac{1}{V} \int d^3 r {\rm Im} \underline{g}_1 (\varepsilon_n, {\bm p}_{\rm F}, {\bm r}) \notag \\ 
&- \frac{1}{2} \frac{N' (\mu_{\rm n})}{N (\mu_{\rm n})} 
\int_{-\tilde{\varepsilon}_{\rm c}}^{\tilde{\varepsilon}_{\rm c}} d \varepsilon 
\frac{\varepsilon}{{\rm e}^{\varepsilon / k_{\rm B} T} + 1} 
\frac{1}{V} \int d^3 r \left[ \frac{N_{{\rm s}0} (\varepsilon, {\bm r})}{N(\mu_{\rm n})} - 2 \right] \notag \\
&+ \frac{(-1)^l}{2} c \frac{N' (\mu_{\rm n})}{N (\mu_{\rm n})} {\rm Tr} \sum_{m=-l}^l \frac{1}{V} \int d^3 r
|\underline{\Delta}_{lm} ({\bm r})|^2 + e \frac{1}{V} \int d^3 r \Phi ({\bm r}). \label{mus} 
\end{align}
Note that Eq. (\ref{mus}) is different from the formula proposed by van der Marel \cite{Marel} and Khomskii and Kusmartsev, \cite{Khomskii92} 
even in the homogeneous $s$-wave pairing case.

\section{Numerical Results \label{sec:III}}

We here perform numerical calculations for the homogeneous and isolated vortex systems of 
clean $s$-wave superconductors with a spherical Fermi surface based on the augmented Eilenberger equations. 
We assume the spin-singlet pairing without spin paramagnetism and 
restrict ourselves to the following forms for the Green's functions and pair potential: 
$\underline{g}_0 = g_0 \underline{\sigma}_0$, $\underline{g}_1 = g_1 \underline{\sigma}_0$,
$\underline{f} = i f \underline{\sigma}_y$, and $\underline{\Delta} = i \Delta \underline{\sigma}_y$, 
where $\underline{\sigma}_y$ is the second Pauli matrix defined by
\begin{align}
\underline{\sigma}_y \equiv 
\begin{bmatrix}
0 & - i  \\
i & 0
\end{bmatrix}. 
\end{align}
The parameters of this system are the coherence length $\xi_0$, 
magnetic penetration depth $\lambda_{0} \equiv \left[\mu_0 N(\mu_{\rm n})e^2 v_{\rm F}^2 \right]^{-1/2}$,
Thomas--Fermi screening length $\lambda_{\rm TF}$, quasiclassical parameter $\delta$, 
and the smearing factor $\eta$ in the advanced and retarded Green functions.
We fixed the parameters to $\lambda_{\rm TF} = 0.01\xi_0$, $\delta = 0.01$, and $\eta = 0.01 \Delta_0$.

\begin{figure}[t]
        \begin{center}
                \includegraphics[width=1.0\linewidth]{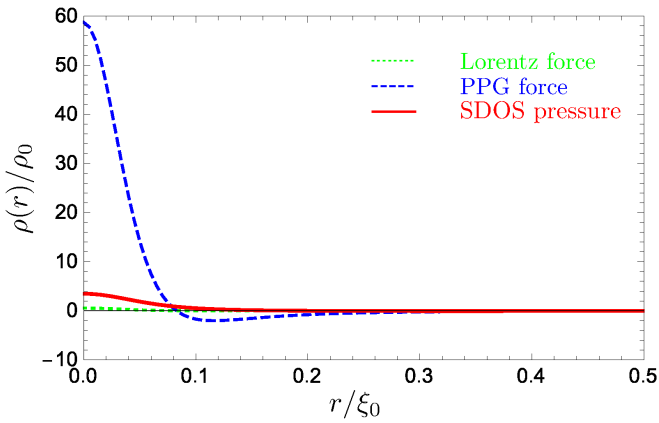}
                \end{center}
\caption{(Color online) Charge redistribution $\rho (r)$ due to the Lorentz force (blue dotted line), 
PPG force (blue dashed line), and SDOS pressure (red solid line)
in units of $\rho_0 \equiv \epsilon_0 \Delta_0 / | e | \xi_0^2$ over $r \le 0.5\xi_0$ for $\lambda_{0} = 5\xi_0$ at $T=0.2T_{\rm c}$.}
\label{fig3}
\end{figure}
\begin{figure}[t]
        \begin{center}
                \includegraphics[width=1.0\linewidth]{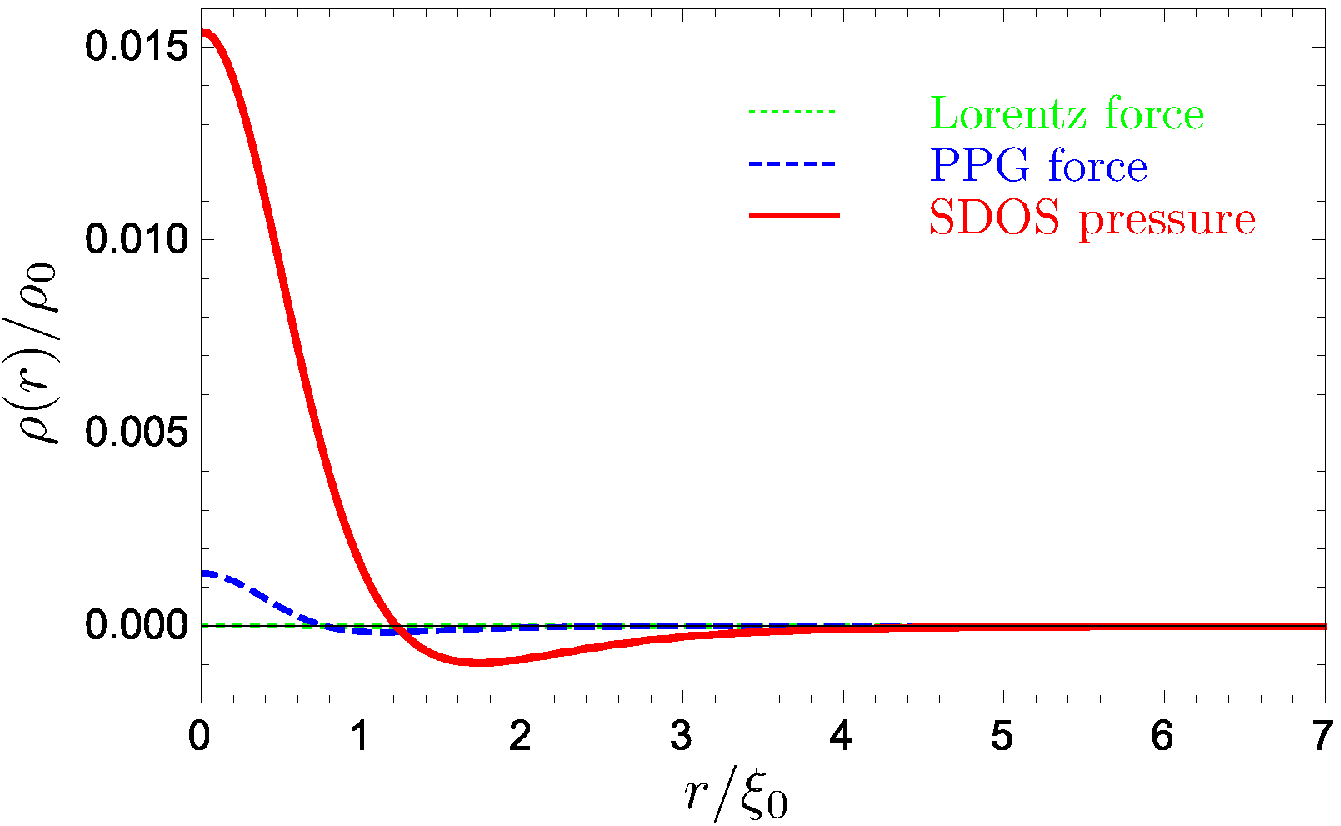}
                \end{center}
\caption{(Color online) Charge redistribution $\rho (r)$ due to the Lorentz force (green dotted line), 
PPG force (blue dashed line), and SDOS pressure (red solid line)
in units of $\rho_0 \equiv \epsilon_0 \Delta_0 / | e | \xi_0^2$ over $r \le 7\xi_0$ for $\lambda_{0} = 5\xi_0$ at $T=0.9T_{\rm c}$.}
\label{fig4}
\end{figure}
\begin{figure}[t]
        \begin{center}
                \includegraphics[width=1.0\linewidth]{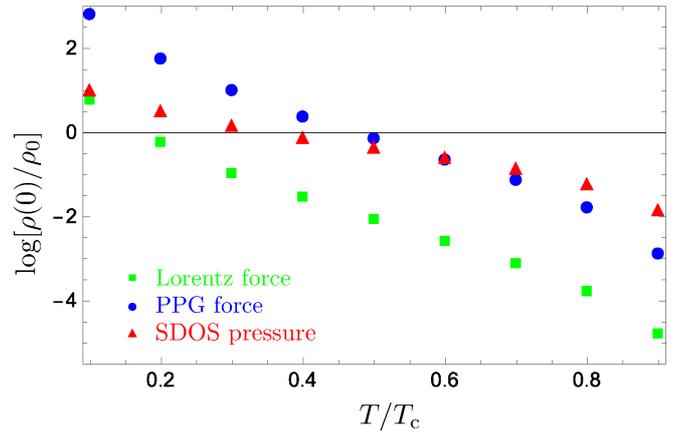}
                \end{center}
\caption{(Color online) Normalized charge density due to the Lorentz force (green square points), 
PPG force (blue circular points), and SDOS pressure (red triangular points)
at the vortex center for $\lambda_{0} = 5\xi_0$ as a function of temperature.}
\label{fig5}
\end{figure}
\begin{figure}[t]
        \begin{center}
                \includegraphics[width=1.0\linewidth]{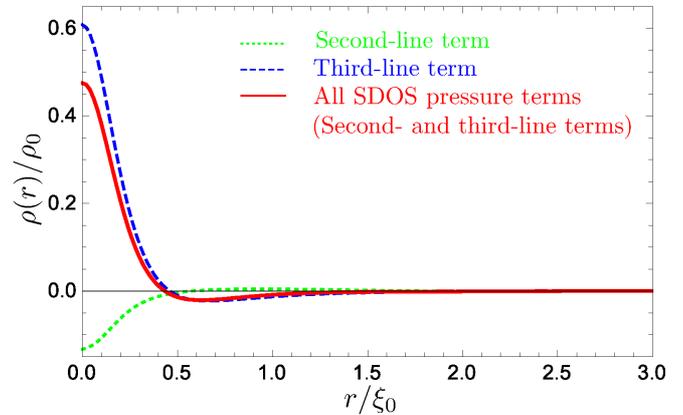}
                \end{center}
\caption{(Color online) Charge redistribution $\rho (r)$ due to the second-line term (green dotted line), 
third-line term (blue dashed line), and all SDOS pressure terms (red solid line) 
in units of $\rho_0 \equiv \epsilon_0 \Delta_0 / | e | \xi_0^2$ over $r \le 3\xi_0$ for $\lambda_{0} = 5\xi_0$ at $T=0.5T_{\rm c}$.}
\label{fig6}
\end{figure}
%

\subsection{Homogeneous system}

First, we present the superconducting DOS and chemical potential in the homogeneous system. 
The Green's functions $g_0$, $g_1$, and $f$ in the homogeneous system are given by 
\begin{align}
g_0 = \frac{\varepsilon_n}{\sqrt{\varepsilon_n^2 + \Delta^2}}, \ \ \ g_1 = 0, \ \ \ 
f = \frac{\Delta}{\sqrt{\varepsilon_n^2 + \Delta^2}}, 
\end{align}
and we can obtain the retarded Green's functions by replacing $\varepsilon_n \to - i \varepsilon + \eta$. 
Substituting them into Eqs. (\ref{Ns3}), (\ref{GapEq}), and (\ref{mus}), 
and solving the gap equation [Eq. (\ref{GapEq})], 
we calculate the superconducting DOS and chemical potential in the homogeneous $s$-wave pairing case.

Figure \ref{fig1} plots the superconducting DOS at $T=0.1T_{\rm c}$.  
We see that Eq. (\ref{Ns3}) can accurately describe the superconducting DOS, 
which has particle-hole asymmetry and connects approximately with the normal DOS at a high energy. 
These behaviors cannot be described by the standard Eilenberger equations without a slope in the DOS. 
Figure \ref{fig2} plots the superconducting and normal chemical potentials as a function of temperature. 
Our obtained superconducting chemical potential is smaller than the normal one, 
and it is consistent with that obtained by van der Marel \cite{Marel,Kato00} and Khomskii and Kusmartsev \cite{Khomskii92}. 
A minimum value of the superconducting chemical potential near $T=0.6T_{\rm c}$ 
arises from the difference in the temperature dependence between the third and fourth terms in Eq. (\ref{mus}) with respect to the slope in the DOS.

\begin{figure}[t]
        \begin{center}
                \includegraphics[width=1.0\linewidth]{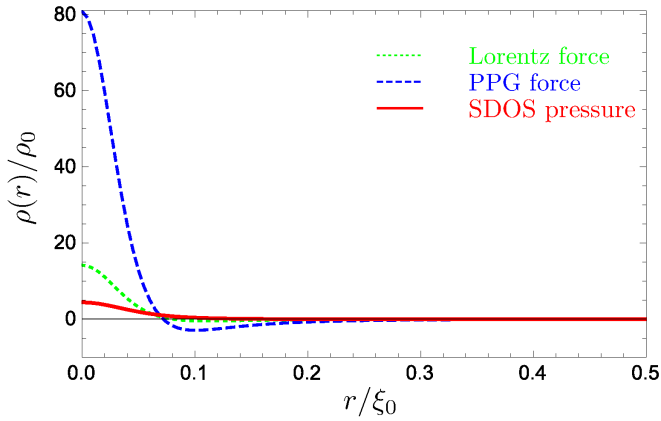}
                \end{center}
\caption{(Color online) Charge redistribution $\rho (r)$ due to the Lorentz force (green dotted line), 
PPG force (blue dashed line), and SDOS pressure (red solid line)
in units of $\rho_0 \equiv \epsilon_0 \Delta_0 / | e | \xi_0^2$ over $r \le 0.5\xi_0$ for $\lambda_{0} = 0.7\xi_0$ at $T=0.2T_{\rm c}$.}
\label{fig7}
\end{figure}
%

\subsection{Vortex-core charging}
We carry out the numerical calculation of vortex-core charging. 
We consider an isolated vortex that is homogeneous along the direction of the magnetic field, 
which is taken parallel to the $z$ axis. 
We also choose a coordinate system where the vortex center is located on the $z$ axis 
and calculate the charge distribution  
at $z = 0$ in a superconductor with thickness from $z = - z_{\rm c}$ to $z = z_{\rm c}$. 
Our numerical procedure is summarized as follows.
We first solve the standard Eilenberger equations self-consistently to obtain 
$(g_0, f, \bar{f}, \Delta, {\bm B})$. 
Although there is a difference depending on whether the Fermi surface is spherical or cylindrical, 
the numerical procedure used to solve the standard Eilenberger equations 
is the same as that described in Sect. 16.3 of Ref. \citen{KitaText}. 
We also obtain $g_0^{\rm R}$ by solving the Riccati-type equation 
for the self-consistent $\Delta$ and ${\bm A}$, replacing $\varepsilon_n \to -i \varepsilon + \eta$ \cite{KitaText}. 
Next, using the standard Runge--Kutta method, we solve the equation for $g_1$, 
which can be obtained from Eq. (\ref{AQCEq}) by procedure described 
in Ref. \citen{Ohuchi} as
\begin{align}
{\bm v}_{\rm F} \cdot {\bm \nabla} g_1 
= - ({\bm v}_{\rm F} \times {\bm B}) \cdot \frac{\partial g_0}{\partial {\bm p}_{\rm F}} 
- \frac{i}{2} {\bm \partial} \Delta^* \cdot \frac{\partial f}{\partial {\bm p}_{\rm F}}
- \frac{i}{2} {\bm \partial} \Delta \cdot \frac{\partial \bar{f}}{\partial {\bm p}_{\rm F}}. \label{g1Eq}
\end{align}
Hence, we can obtain $g_1$ from the solution for the standard Eilenberger equations.
Substituting ${\rm Im} g_1$ and ${\rm Re} g_0^{\rm R}$ obtained by the above procedures into Eq. (\ref{EEq}) 
and solving Eq. (\ref{EEq}) numerically, we finally obtain the electric field and charge density.

Figures \ref{fig3} and \ref{fig4} plot the charge density due to the three forces 
in the core region for $\lambda_{0} = 5\xi_0$ at $T=0.2T_{\rm c}$ and $T=0.9T_{\rm c}$, respectively, 
and Fig. \ref{fig5} plots the logarithm of the charge density at the vortex center for $\lambda_{0} = 5\xi_0$ as a function of temperature. 
It is shown that the charge densities due to the three forces at the vortex center increase as the temperature decreases from $T=T_{\rm c}$. 
Since the charge densities due to the Lorentz and PPG forces are proportional to ${\bm B} $ and ${\bm \partial} \Delta$, 
the enhancement of the charge accumulation by the Lorentz and PPG forces 
can be attributed to the decrease in the London penetration depth and to the core shrinkage due to the Kramer--Pesch effect, \cite{Ohuchi}
respectively, that is, the core size of the vortex narrows as the temperature decreases from $T=T_{\rm c}$ 
and approaches zero as $T \to 0$ \cite{Kramer,KitaText}. 
On the other hand, the charge density due to the SDOS pressure is proportional to $| \Delta |$ 
because the core charge due to the second-line term in Eq. (\ref{rho2}) is smaller than 
that due to the third-line term in Eq. (\ref{rho2}) within this parameter range, but these signs are opposite to each other. 
As an example, Fig. \ref{fig6} shows a comparison between the second- and third-line terms in Eq. (\ref{rho2}) for $\lambda_{0} = 5\xi_0$ at $T=0.5T_{\rm c}$. 
Thus, the charge accumulation in the core region caused by the SDOS pressure is enhanced owing to the increase in the magnitude of $\Delta$ 
as the temperature decreases, 
and we can neglect the second-line term in Eq. (\ref{rho2}) to obtain a rough estimate of the vortex-core charge.
We also observe that the core charge due to the Lorentz force is negligible compared with the charges due to other forces,
and the core charge due to the SDOS pressure is larger near $T = T_{\rm c}$  
and smaller near $T = 0$ than that due to the PPG force. 
The core charge due to the PPG force, which is proportional to ${\bm \partial} \Delta$, is strongly affected by the Kramer--Pesch effect.
Thus, the core charge due to the PPG force is more enhanced by the Kramer--Pesch effect near $T = 0$. 
We also see that the slope in the pair potential reaches zero faster than the magnitude of the pair potential as $T \to T_{\rm c}$, 
so the core charge due to the PPG force is smaller than that due to the SDOS pressure near $T = T_{\rm c}$.

\begin{figure}[t]
        \begin{center}
                \includegraphics[width=1.0\linewidth]{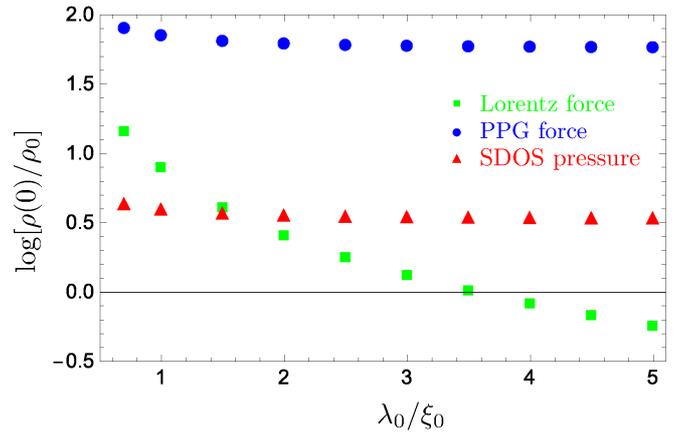}
                \end{center}
\caption{(Color online) Normalized charge density due to the Lorentz force (green square points), 
PPG force (blue circular points), and SDOS pressure (red triangular points)
at the vortex center for $\lambda_{0} = 5\xi_0$ as a function of $\lambda_0$ calculated for $T=0.2T_{\rm c}$.}
\label{fig8}
\end{figure}

Figure \ref{fig7} plots the charge density due to the three forces 
in the core region for $\lambda_{0} = 0.7\xi_0$ at $T=0.2T_{\rm c}$,
and Fig. \ref{fig8} plots the logarithm of the charge density due to the three forces at the vortex center for $T=0.2T_{\rm c}$ as a function of $\lambda_0$. 
The magnetic field in the core region is enhanced as $\lambda_0$ decreases, 
so the core charge due to the Lorentz force is negligible compared with that due to the other forces for $\lambda_0 \gtrsim 2 \xi_0$, 
but becomes substantial over the value of that due to the SDOS pressure when $\lambda_0 \sim \xi_0$ \cite{Ohuchi}. 
On the other hand, since the shape of the pair potential hardly changes with $\lambda_0$, 
the charge accumulation in the core induced by the SDOS pressure hardly changes with $\lambda_0$ compared with that induced by the Lorentz force 
and has the same behavior as that induced by the PPG force.

%

\section{Summary \label{sec:IV}} 

We derived augmented Eilenberger equations incorporating the Lorentz force, PPG force, and SDOS pressure.
Using them, we calculated the DOS and chemical potential in homogeneous $s$-wave superconductors with a spherical Fermi surface. 
The standard Eilenberger equations can include the effect of the anisotropic Fermi surface more easily than the BdG equations. 
However, the superconducting DOS obtained from the equations without the slope in the DOS 
always satisfies the particle-hole symmetry regardless of the anisotropic Fermi surface. 
On the other hand, our numerical result has particle-hole asymmetry and connects approximately with the normal DOS. 
Thus, our method may more accurately reproduce the superconducting DOS in a variety of materials 
measured in STS experiments \cite{Renner,Aprile,Nishida,Kugler,Fischer,Hanaguri10,Hanaguri12}.
We also derived the formula for the chemical potential difference between the normal and superconducting states arising 
from the Lorentz force, PPG force, and SDOS pressure using the augmented Eilenberger equations.

We compared which forces dominantly contribute to charging in an isolated vortex of an $s$-wave superconductor 
with a spherical Fermi surface using the augmented Eilenberger equations. 
We observed that 
when the London penetration depth is much larger than the coherence length, 
the contribution of the Lorentz force to the vortex-core charge is negligibly small compared with that of the other forces, 
the contribution of the SDOS pressure becomes dominant near the transition temperature, 
and the contribution of the PPG force is so large that the other forces are negligible near absolute zero temperature. 
We also found that when the London penetration depth is about the same as the coherence length, 
the contribution of the Lorentz force to the core charge becomes substantial over the value of that of the SDOS pressure, 
but smaller than that of the PPG force. 
On the other hand, our previous works have shown that 
the charge accumulation by the Lorentz force in the core is strongly enhanced compared with that of an isolated vortex 
as the external magnetic field increases from a lower critical field \cite{Kohno16,Kohno17}. 
Thus, the contribution of the Lorentz force to the core charge may become dominant over the contributions of other forces in finite magnetic fields.
Hence, we should study the magnetic-field dependence of the vortex-core charging due to the three forces 
by solving these augmented Eilenberger equations with three forces for the vortex lattice system. 
Moreover, in the flux-flow state, we need to consider the effect of the Lorentz force on not only suppercurrent but also normal current in the core region. 
The contribution of the Lorentz force to the flux-flow Hall effect may therefore be larger than the contributions of the other forces. 
Thus, because the force that dominantly contributes to quantities 
may vary with the parameters of the materials, the temperature, the external field, and the system, 
we need to consider all three forces to study charging and transport phenomena 
such as the flux-flow Hall effect in type-II superconductors.

\begin{acknowledgment}
The computation in this work was carried out using the facilities of the Supercomputer Center, 
the Institute for Solid State Physics, the University of Tokyo. 
\end{acknowledgment}

\appendix 
\section{Kinetic-Energy Terms in the Wigner Representation \label{AppA}}

Let us simplify the kinetic-energy terms of the Gor'kov equation [Eq. (\ref{Gor'kovEq})] 
in the Wigner representation [Eq. (\ref{GIWT2})]. 

We introduce the functions
\begin{align}
\mathcal{E}_1 (u) &\equiv \int_0^1 d \eta {\rm e}^{\eta u} = \frac{{\rm e}^{u} - 1}{u} = \sum_{n=1}^\infty \frac{u^{n-1}}{n!}, \\ 
\mathcal{E}_2 (u) &\equiv \int_0^1 d \eta \int_0^\eta d \zeta {\rm e}^{\zeta u} = \frac{{\rm e}^{u} - 1 - u}{u^2} 
= \sum_{n=2}^\infty \frac{u^{n-2}}{n!}, 
\end{align}
with which we can express the basic phase factors that appear in Eq. (\ref{GIWT2}) as
\begin{subequations}
\begin{align}
I ({\bm r}_1, {\bm r}_{12}) 
&= \frac{e}{\hbar} \mathcal{E}_1 \left( \frac{\bar{\bm r}_{12}}{2} \cdot \frac{\partial}{\partial {\bm r}_{12}} \right) 
{\bf A} ({\bf r}_{12}) \cdot \frac{\bar{\bm r}_{12}}{2}, \\
I ({\bm r}_{12}, {\bm r}_2) 
&= \frac{e}{\hbar} \mathcal{E}_1 \left( - \frac{\bar{\bm r}_{12}}{2} \cdot \frac{\partial}{\partial {\bm r}_{12}} \right) 
{\bf A} ({\bf r}_{12}) \cdot \frac{\bar{\bm r}_{12}}{2}. 
\end{align} \label{I2}
\end{subequations}
Using $\partial / \partial {\bm r}_1 = \partial / \partial \bar{\bm r}_{12} + (1 / 2) \partial / \partial {\bm r}_{12}$
and Eq. (\ref{I2}), we can transform 
$(\partial / \partial {\bm r}_1) I ({\bm r}_1, {\bm r}_{12})$ and $(\partial / \partial {\bm r}_1) I ({\bm r}_{12}, {\bm r}_2)$ as
\begin{subequations}
\begin{align}
&\frac{\partial}{\partial {\bm r}_1} I ({\bm r}_1, {\bm r}_{12}) 
= \frac{e}{\hbar} {\bm A} ({\bm r}_1) - \frac{e}{2 \hbar} {\bm A} ({\bm r}_{12}) \notag \\
& \ \ \ \ \ - \frac{e}{4 \hbar} \left[ 2 \mathcal{E}_1 \left( \frac{\bar{\bm r}_{12}}{2} \cdot \frac{\partial}{\partial {\bm r}_{12}} \right) 
- \mathcal{E}_2 \left( \frac{\bar{\bm r}_{12}}{2} \cdot \frac{\partial}{\partial {\bm r}_{12}} \right) \right] 
{\bm B} ({\bm r}_{12}) \times \bar{\bm r}_{12}, \\
&\frac{\partial}{\partial {\bm r}_1} I ({\bm r}_{12}, {\bm r}_2) 
= \frac{e}{2 \hbar} {\bm A} ({\bm r}_{12}) 
- \frac{e}{4 \hbar} \mathcal{E}_2 \left( -\frac{\bar{\bm r}_{12}}{2} \cdot \frac{\partial}{\partial {\bm r}_{12}} \right) 
{\bm B} ({\bm r}_{12}) \times \bar{\bm r}_{12}. 
\end{align} \label{dr1I}
\end{subequations} 
Now, we focus on the kinetic-energy terms in Eq. (\ref{Gor'kovEq}) given by
\begin{align}
&
\begin{bmatrix}
\hat{\mathcal{K}}_1 \underline{\sigma}_0  & \underline{0} \\
\underline{0} & - \hat{\mathcal{K}}_1^* \underline{\sigma}_0
\end{bmatrix}
\hat{G} ({\bm r}_1, {\bm r}_2 ; \varepsilon_n) \notag \\
& \ \ \ \ \ = 
\begin{bmatrix}
\hat{\mathcal{K}}_1 \underline{G} ({\bm r}_1, {\bm r}_2 ; \varepsilon_n) 
& \hat{\mathcal{K}}_1 \underline{F} ({\bm r}_1, {\bm r}_2 ; \varepsilon_n) \\
\hat{\mathcal{K}}_1^* \underline{F}^* ({\bm r}_1, {\bm r}_2 ; \varepsilon_n) 
& \hat{\mathcal{K}}_1^* \underline{G}^* ({\bm r}_1, {\bm r}_2 ; \varepsilon_n)
\end{bmatrix}.
\end{align}
Substituting Eq. (\ref{GIWT2}) and using 
$\partial / \partial {\bm r}_1 = \partial / \partial \bar{\bm r}_{12} + (1 / 2) \partial / \partial {\bm r}_{12}$
and Eq. (\ref{dr1I}), we can transform each submatrix on the right-hand side as
\begin{subequations}
\label{KGF}
\begin{align}
&\hat{\mathcal{K}}_1 \underline{G} ({\bm r}_1, {\bm r}_2 ; \varepsilon_n) 
\approx {\rm e}^{i I ({\bm r}_1, {\bm r}_{12}) + i I ({\bm r}_{12}, {\bm r}_2)} 
\int \frac{d^3 p}{(2 \pi \hbar)^3} {\rm e}^{i {\bm p} \cdot \bar{{\bf r}}_{12} / \hbar} \notag \\
&\ \ \ \times \Bigg\{ \frac{p^2}{2m} + e \Phi ({\bm r}_{12}) - \mu
- \frac{i \hbar}{2} \frac{\bm p}{m} \cdot \frac{\partial}{\partial {\bm r}_{12}} 
- \frac{\hbar^2}{8m} \frac{\partial^2}{\partial {\bm r}_{12}^2} \notag \\
& \ \ \ - \frac{i \hbar}{2} e \frac{\bm p}{m} \cdot \left[{\bm B} ({\bm r}_{12}) \times \frac{\partial}{\partial {\bm p}} \right]
- \frac{i \hbar}{2} e {\bm E} ({\bm r}_{12}) \cdot \frac{\partial}{\partial {\bm p}} \Bigg\} 
\underline{G} (\varepsilon_n, {\bm p}, {\bm r}_{12}), 
\end{align}
\begin{align}
&\hat{\mathcal{K}}_1 \underline{F} ({\bm r}_1, {\bm r}_2 ; \varepsilon_n) 
\approx {\rm e}^{i I ({\bm r}_1, {\bm r}_{12}) - i I ({\bm r}_{12}, {\bm r}_2)} 
\int \frac{d^3 p}{(2 \pi \hbar)^3} {\rm e}^{i {\bm p} \cdot \bar{{\bf r}}_{12} / \hbar} \notag \\
& \ \ \ \times \Bigg\{ \frac{p^2}{2m} + e \Phi ({\bm r}_{12}) - \mu
- \frac{i \hbar}{2} \frac{\bm p}{m} \cdot \left[ \frac{\partial}{\partial {\bm r}_{12}} - i \frac{2e}{\hbar} {\bm A} ({\bm r}_{12}) \right] 
\notag \\
& \ \ \ - \frac{\hbar^2}{8m} \left[ \frac{\partial}{\partial {\bm r}_{12}} - i \frac{2e}{\hbar} {\bm A} ({\bm r}_{12}) \right]^2
- \frac{i \hbar}{4} e \frac{\bm p}{m} \cdot \left[{\bm B} ({\bm r}_{12}) \times \frac{\partial}{\partial {\bm p}} \right] \notag \\
& \ \ \ - \frac{i \hbar}{2} e {\bm E} ({\bm r}_{12}) \cdot \frac{\partial}{\partial {\bm p}} \Bigg\} 
\underline{F} (\varepsilon_n, {\bm p}, {\bm r}_{12}), 
\end{align}
\begin{align}
&\hat{\mathcal{K}}_1^* \underline{F}^* ({\bm r}_1, {\bm r}_2 ; \varepsilon_n) 
\approx {\rm e}^{- i I ({\bm r}_1, {\bm r}_{12}) + i I ({\bm r}_{12}, {\bm r}_2)} 
\int \frac{d^3 p}{(2 \pi \hbar)^3} {\rm e}^{i {\bm p} \cdot \bar{{\bf r}}_{12} / \hbar} \notag \\
& \ \ \ \Bigg\{ \frac{p^2}{2m} + e \Phi ({\bm r}_{12}) - \mu
- \frac{i \hbar}{2} \frac{\bm p}{m} \cdot 
\left[ \frac{\partial}{\partial {\bm r}_{12}} + i \frac{2e}{\hbar} {\bm A} ({\bm r}_{12}) \right] \notag \\
& \ \ \ - \frac{\hbar^2}{8m} \left[ \frac{\partial}{\partial {\bm r}_{12}} + i \frac{2e}{\hbar} {\bm A} ({\bm r}_{12}) \right]^2
+ \frac{i \hbar}{4} e \frac{\bm p}{m} \cdot \left[{\bm B} ({\bm r}_{12}) \times \frac{\partial}{\partial {\bm p}} \right] \notag \\
& \ \ \ - \frac{i \hbar}{2} e {\bm E} ({\bm r}_{12}) \cdot \frac{\partial}{\partial {\bm p}} \Bigg\} 
\underline{F}^* (\varepsilon_n, - {\bm p}, {\bm r}_{12}), 
\end{align}
\begin{align}
&\hat{\mathcal{K}}_1^* \underline{G}^* ({\bm r}_1, {\bm r}_2 ; \varepsilon_n) 
\approx {\rm e}^{- i I ({\bm r}_1, {\bm r}_{12}) - i I ({\bm r}_{12}, {\bm r}_2)} 
\int \frac{d^3 p}{(2 \pi \hbar)^3} {\rm e}^{i {\bm p} \cdot \bar{{\bf r}}_{12} / \hbar} \notag \\
& \ \ \ \times \Bigg\{ \frac{p^2}{2m} + e \Phi ({\bm r}_{12}) - \mu
- \frac{i \hbar}{2} \frac{\bm p}{m} \cdot \frac{\partial}{\partial {\bm r}_{12}} 
- \frac{\hbar^2}{8m} \frac{\partial^2}{\partial {\bm r}_{12}^2} \notag \\
& \ \ \ + \frac{i \hbar}{2} e \frac{\bm p}{m} \cdot \left[{\bm B} ({\bm r}_{12}) \times \frac{\partial}{\partial {\bm p}} \right]
- \frac{i \hbar}{2} e {\bm E} ({\bm r}_{12}) \cdot \frac{\partial}{\partial {\bm p}} \Bigg\} 
\underline{G}^* (\varepsilon_n, - {\bm p}, {\bm r}_{12}). 
\end{align} \label{KineticEnegyTerms}
\end{subequations}
The following approximations have been adopted in deriving Eq.\ (\ref{KGF}). 
(i) We have neglected spatial derivatives of both ${\bm E}$ and ${\bm B}$, which amounts to setting
 $\mathcal{E}_1 \to 1$ and $\mathcal{E}_2 \to 1/2$. 
(ii) We have also neglected the second-order terms in ${\bm \partial}_{{\bm r}_{12}}$, ${\bm E}$, and ${\bm B}$ 
except that of ${\bm \partial}_{{\bm r}_{12}}^2$. 
(iii) We have expanded $\Phi$ around ${\bm r}_{12}$ up to the first order in $\bar{\bm r}_{12}$ as 
$\Phi ({\bm r}_1) \approx \Phi ({\bm r}_{12}) - {\bm E} ({\bm r}_{12}) \cdot \bar{\bm r}_{12} / 2$. 
By following these procedures, 
we obtain the kinetic-energy terms of the Gor'kov equation [Eq. (\ref{Gor'kovEq})] in the Wigner representation as 
\begin{align}
&\int d^3 \bar{r}_{12} {\rm e}^{- i {\bm p} \cdot \bar{\bm r}_{12} / \hbar} \notag \\
& \ \ \ \ \ \times \hat{\Gamma} ({\bm r}_{12}, {\bm r}_1) 
\begin{bmatrix}
\hat{\mathcal{K}}_1 \underline{\sigma}_0  & \underline{0} \\
\underline{0} & - \hat{\mathcal{K}}_1^* \underline{\sigma}_0
\end{bmatrix}
\hat{G} ({\bm r}_1, {\bm r}_2 ; \varepsilon_n) 
\hat{\Gamma} ({\bm r}_2, {\bm r}_{12}) \notag \\
&= 
\Bigg[ \frac{p^2}{2 m} + e \Phi ({\bm r}_{12}) - \mu 
- \frac{i \hbar}{2} \frac{\bm p}{m} \cdot {\bm \partial}_{12} - \frac{\hbar^2}{8m} {\bm \partial}_{12}^2 \notag \\
& \ \ \ \ \ - \frac{i \hbar}{2} e {\bm E} ({\bm r}_{12}) \cdot \frac{\partial}{\partial {\bm p}} \Bigg] 
\hat{\tau}_3 \hat{G} (\varepsilon_n, {\bm p}, {\bm r}_{12}) \notag \\
& - \frac{i \hbar}{8} e \frac{\bm p}{m} \cdot \left[ {\bm B} ({\bm r}_{12}) \times \frac{\partial}{\partial {\bm p}} \right] 
\left[ 3 \hat{G} (\varepsilon_n, {\bm p}, {\bm r}_{12}) 
+ \hat{\tau}_3 \hat{G} (\varepsilon_n, {\bm p}, {\bm r}_{12}) \hat{\tau}_3 \right]. \label{KineticEnegyTerms2}
\end{align}
%

\section{Self-Energy Terms in the Wigner Representation \label{AppB}}

\begin{figure}[t]
        \begin{center}
                \includegraphics[width=1.0\linewidth]{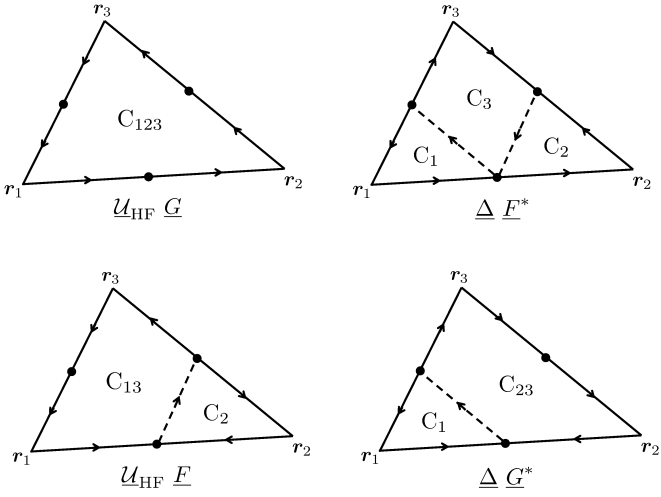}
                \end{center}
\caption{Paths of the phase integrals.}
\label{figB1}
\end{figure}

We consider the self-energy terms in Eq. (\ref{Gor'kovEq}). 
Let us substitute Eqs. (\ref{GIWT2}) and (\ref{UIWT2}) into Eq. (\ref{Gor'kovEq}) and express the resulting expression as
\begin{align}
&\int d^3 r_3 \hat{\cal U}_{\rm BdG} ({\bm r}_1, {\bm r}_3 ) \hat{G} ({\bm r}_3, {\bm r}_2 ; \varepsilon_n)= \notag \\
& 
\begin{bmatrix}
\underline{J} ({\bm r}_1, {\bm r}_2 ; \varepsilon_n) - \underline{K} ({\bm r}_1, {\bm r}_2 ; \varepsilon_n)  
\ \ \ \ \underline{L} ({\bm r}_1, {\bm r}_2 ; \varepsilon_n) - \underline{M} ({\bm r}_1, {\bm r}_2 ; \varepsilon_n) \\
\underline{L}^* ({\bm r}_1, {\bm r}_2 ; \varepsilon_n) - \underline{M}^* ({\bm r}_1, {\bm r}_2 ; \varepsilon_n) 
\ \ \underline{J}^* ({\bm r}_1, {\bm r}_2 ; \varepsilon_n) - \underline{K}^* ({\bm r}_1, {\bm r}_2 ; \varepsilon_n)
\end{bmatrix}, \label{SelfEnegyTerms}%
\end{align}
where matrices $\underline{J}({\bm r}_1, {\bm r}_2 ; \varepsilon_n)$, $\underline{K}({\bm r}_1, {\bm r}_2 ; \varepsilon_n)$, 
$\underline{L}({\bm r}_1, {\bm r}_2 ; \varepsilon_n)$, and $\underline{M}({\bm r}_1, {\bm r}_2 ; \varepsilon_n)$ are defined by 
\begin{subequations}
\begin{align}
&\underline{J} ({\bm r}_1, {\bm r}_2 ; \varepsilon_n) 
\equiv \int d^3 r_3 \underline{\cal U}_{\rm HF} ({\bm r}_1, {\bm r}_3) \underline{G} ({\bm r}_3, {\bm r}_2; \varepsilon_n), 
\label{J} \\
&\underline{K} ({\bm r}_1, {\bm r}_2 ; \varepsilon_n) 
\equiv \int d^3 r_3 \underline{\Delta} ({\bm r}_1, {\bm r}_3) \underline{F}^* ({\bm r}_3, {\bm r}_2; \varepsilon_n), \label{K} \\
&\underline{L} ({\bm r}_1, {\bm r}_2 ; \varepsilon_n) 
\equiv \int d^3 r_3 \underline{\cal U}_{\rm HF} ({\bm r}_1, {\bm r}_3) \underline{F} ({\bm r}_3, {\bm r}_2; \varepsilon_n), 
\label{L} \\
&\underline{M} ({\bm r}_1, {\bm r}_2 ; \varepsilon_n) 
\equiv \int d^3 r_3 \underline{\Delta} ({\bm r}_1, {\bm r}_3) \underline{G}^* ({\bm r}_3, {\bm r}_2; \varepsilon_n). \label{M}
\end{align}\label{SelfEnegyTerms2}%
\end{subequations}

We first focus on Eq. (\ref{J}). 
Substituting Eqs. (\ref{GIWT2}) and (\ref{UIWT2}) into Eq. (\ref{J}), 
we obtain the matrix $\underline{J} ({\bm r}_1, {\bm r}_2 ; \varepsilon_n)$ as
\begin{align}
&\underline{J} ({\bm r}_1, {\bm r}_2 ; \varepsilon_n) 
= {\rm e}^{i I ({\bm r}_1, {\bm r}_{12}) + i I ({\bm r}_{12}, {\bm r}_2)} 
\int \frac{d^3 p}{(2 \pi \hbar)^3} \int \frac{d^3 p'}{(2 \pi \hbar)^3} 
\int d^3 r_3 \notag \\
&\ \ \ \ \ \times {\rm e}^{i \phi_{123} + i {\bm p} \cdot \bar{\bm r}_{13} / \hbar + i {\bm p}' \cdot \bar{\bm r}_{32} / \hbar } 
\underline{\cal U}_{\rm HF} ({\bm p}, {\bm r}_{13}) \underline{G} (\varepsilon_n, {\bm p}', {\bm r}_{32}), 
\end{align}
where the phase integral $\phi_{123}$ is defined by 
\begin{align}
\phi_{123} \equiv \frac{e}{\hbar} \oint_{C_{123}} {\bm A} ({\bm s}) \cdot d {\bm s},  
\end{align}
with the integral path $C_{123}$ given in Fig. \ref{figB1}. 
Using the Stokes theorem, approximating ${\bm B} ({\bm r}) \approx {\bm B} ({\bm r}_{12})$ and 
noting Fig. \ref{figB1}, 
the phase integral $\phi_{123}$ is given by 
\begin{align}
\phi_{123} 
= \frac{e}{\hbar} \int_{S_{123}} {\bm B} ({\bm r}) \cdot d {\bm S}
\approx \frac{e}{2 \hbar} {\bm B} ({\bm r}_{12}) \cdot (\bar{\bm r}_{32} \times \bar{\bm r}_{13}). \label{phi123}
\end{align}
By the same procedure as the standard Wigner transformation, \cite{KitaReview}
we obtain the matrix $\underline{J} ({\bm r}_1, {\bm r}_2 ; \varepsilon_n)$ 
with $\underline{\cal U}_{\rm HF} ({\bm p}, {\bm r}_{12})$ and $\underline{G} (\varepsilon_n, {\bm p}, {\bm r}_{12})$ 
in the Wigner representation as 
\begin{align}
&\underline{J} ({\bm r}_1, {\bm r}_2 ; \varepsilon_n) 
\approx {\rm e}^{i I ({\bm r}_1, {\bm r}_{12}) + i I ({\bm r}_{12}, {\bm r}_2)} 
\int \frac{d^3 p}{(2 \pi \hbar)^3} {\rm e}^{i {\bm p} \cdot \bar{\bm r}_{12} / \hbar} 
\underline{\cal U}_{\rm HF} ({\bm p}, {\bm r}_{12}) \notag \\
& \ \ \ \ \ \times 
{\rm e}^{(i \hbar / 2) e {\bm B} ({\bm r}_{12}) \cdot 
(\overleftarrow{\bm \partial}_{\bm p} \times \overrightarrow{\bm \partial}_{\bm p})} 
{\rm e}^{(i \hbar / 2) \overleftarrow{\bm \partial}_{12} \cdot \overrightarrow{\bm \partial}_{\bm p} 
- (i \hbar / 2) \overleftarrow{\bm \partial}_{\bm p} \cdot \overrightarrow{\bm \partial}_{12}} 
 \underline{G} (\varepsilon_n, {\bm p}, {\bm r}_{12}), \label{J2}
\end{align}
where the left (right) arrow on each differential operator denotes that it acts on the left potential (right Green's function).

We next consider Eq. (\ref{K}). 
Let us substitute Eqs. (\ref{GIWT2}) and (\ref{UIWT2}) into Eq. (\ref{M}). 
Then, we can express $\underline{K} ({\bm r}_1, {\bm r}_2 ; \varepsilon_n)$ as 
\begin{align}
&\underline{K} ({\bm r}_1, {\bm r}_2 ; \varepsilon_n) 
= {\rm e}^{i I ({\bm r}_1, {\bm r}_{12}) + i I ({\bm r}_{12}, {\bm r}_2)} 
\int \frac{d^3 p}{(2 \pi \hbar)^3} \int \frac{d^3 p'}{(2 \pi \hbar)^3} 
\int d^3 r_3 \notag \\
& \ \ \ \ \ \times {\rm e}^{i (\phi_1+\phi_2+\phi_3) 
- 2 i I ({\bm r}_{13}, {\bm r}_{12}) - 2 i I ({\bm r}_{12}, {\bm r}_{32})
+ i {\bm p} \cdot \bar{\bm r}_{13} / \hbar + i {\bm p}' \cdot \bar{\bm r}_{32} / \hbar } \notag \\
& \ \ \ \ \ \times \underline{\Delta} ({\bm p}, {\bm r}_{13}) \underline{F}^* (\varepsilon_n, - {\bm p}', {\bm r}_{32}), 
\end{align}
where the phase integrals $\phi_1+\phi_2+\phi_3$ are defined by 
\begin{align}
&\phi_1+\phi_2+\phi_3 \notag \\ 
&\equiv \frac{e}{\hbar} \oint_{C_1} {\bm A} ({\bm s}) \cdot d {\bm s}
+ \frac{e}{\hbar} \oint_{C_2} {\bm A} ({\bm s}) \cdot d {\bm s}
+ \frac{e}{\hbar} \oint_{C_3} {\bm A} ({\bm s}) \cdot d {\bm s}. 
\end{align}
Noting the integral paths $C_1$, $C_2$, and $C_3$ given in Fig. \ref{figB1}, 
we see that $\phi_1+\phi_2+\phi_3 = 0$. 
Thus, the matrix $\underline{K} ({\bm r}_1, {\bm r}_2 ; \varepsilon_n)$ 
with $\underline{\Delta} ({\bm p}, {\bm r}_{12})$ and $\underline{F}^* (\varepsilon_n, {\bm p}, {\bm r}_{12})$ 
in the Wigner representation is given as 
\begin{align}
&\underline{K} ({\bm r}_1, {\bm r}_2 ; \varepsilon_n) 
\approx {\rm e}^{i I ({\bm r}_1, {\bm r}_{12}) + i I ({\bm r}_{12}, {\bm r}_2)} 
\int \frac{d^3 p}{(2 \pi \hbar)^3} {\rm e}^{i {\bm p} \cdot \bar{\bm r}_{12} / \hbar} \notag \\
& \ \ \ \times \underline{\Delta} ({\bm p}, {\bm r}_{12}) 
{\rm e}^{(i \hbar / 2) \overleftarrow{\bm \partial}_{12} \cdot \overrightarrow{\bm \partial}_{\bm p} 
- (i \hbar / 2) \overleftarrow{\bm \partial}_{\bm p} \cdot \overrightarrow{\bm \partial}_{12}}
\underline{F}^* (\varepsilon_n, - {\bm p}, {\bm r}_{12}). \label{K2}
\end{align}

Finally, we calculate $\underline{L} ({\bm r}_1, {\bm r}_2 ; \varepsilon_n)$ 
and $\underline{M} ({\bm r}_1, {\bm r}_2 ; \varepsilon_n)$. 
Substituting Eqs. (\ref{GIWT2}) and (\ref{UIWT2}) into Eqs. (\ref{L}) and (\ref{M}), 
the matrices $\underline{L} ({\bm r}_1, {\bm r}_2 ; \varepsilon_n)$ 
and $\underline{M} ({\bm r}_1, {\bm r}_2 ; \varepsilon_n)$ are given as
\begin{align}
&\underline{L} ({\bm r}_1, {\bm r}_2 ; \varepsilon_n) 
= {\rm e}^{i I ({\bm r}_1, {\bm r}_{12}) - i I ({\bm r}_{12}, {\bm r}_2)} 
\int \frac{d^3 p}{(2 \pi \hbar)^3} \int \frac{d^3 p'}{(2 \pi \hbar)^3} 
\int d^3 r_3 \notag \\
& \ \ \ \times {\rm e}^{i (\phi_{13}+\phi_2) 
- 2 i I ({\bm r}_{32}, {\bm r}_{12}) 
+ i {\bm p} \cdot \bar{\bm r}_{13} / \hbar + i {\bm p}' \cdot \bar{\bm r}_{32} / \hbar } 
\underline{\cal U}_{\rm HF} ({\bm p}, {\bm r}_{13}) \underline{F} (\varepsilon_n, {\bm p}', {\bm r}_{32}), 
\end{align}
\begin{align}
&\underline{M} ({\bm r}_1, {\bm r}_2 ; \varepsilon_n) 
= {\rm e}^{i I ({\bm r}_1, {\bm r}_{12}) - i I ({\bm r}_{12}, {\bm r}_2)} 
\int \frac{d^3 p}{(2 \pi \hbar)^3} \int \frac{d^3 p'}{(2 \pi \hbar)^3} 
\int d^3 r_3 \notag \\
& \ \ \ \times {\rm e}^{i (\phi_1+\phi_{23}) 
- 2 i I ({\bm r}_{13}, {\bm r}_{12}) 
+ i {\bm p} \cdot \bar{\bm r}_{13} / \hbar + i {\bm p}' \cdot \bar{\bm r}_{32} / \hbar } 
\underline{\Delta} ({\bm p}, {\bm r}_{13}) \underline{G}^* (\varepsilon_n, - {\bm p}', {\bm r}_{32}), 
\end{align} 
with the phase integrals $\phi_{13}+\phi_2$ and $\phi_1+\phi_{23}$ defined by 
\begin{align}
&\phi_{13}+\phi_2 \equiv \frac{e}{\hbar} \oint_{C_{13}} {\bm A} ({\bm s}) \cdot d {\bm s}
+ \frac{e}{\hbar} \oint_{C_2} {\bm A} ({\bm s}) \cdot d {\bm s}, \\
&\phi_1+\phi_{23} \equiv \frac{e}{\hbar} \oint_{C_1} {\bm A} ({\bm s}) \cdot d {\bm s}
+ \frac{e}{\hbar} \oint_{C_{23}} {\bm A} ({\bm s}) \cdot d {\bm s}.
\end{align}
By the same calculation as for Eq. (\ref{phi123}), we can carry out the integration of the phase integrals
$\phi_{13}+\phi_2$ and $\phi_1+\phi_{23}$ as
\begin{align}
& \phi_{13} + \phi_2 \approx \frac{e}{4 \hbar} {\bm B} ({\bm r}_{12}) \cdot (\bar{\bm r}_{32} \times \bar{\bm r}_{13}), \\
&\phi_{23} + \phi_1 \approx - \frac{e}{4 \hbar} {\bm B} ({\bm r}_{12}) \cdot (\bar{\bm r}_{32} \times \bar{\bm r}_{13}).
\end{align}
Thus, we obtain the matrices $\underline{L} ({\bm r}_1, {\bm r}_2 ; \varepsilon_n)$ and 
$\underline{M} ({\bm r}_1, {\bm r}_2 ; \varepsilon_n)$
with the potentials and Green's functions in the Wigner representation as
\begin{align}
&\underline{L} ({\bm r}_1, {\bm r}_2 ; \varepsilon_n) 
\approx {\rm e}^{i I ({\bm r}_1, {\bm r}_{12}) - i I ({\bm r}_{12}, {\bm r}_2)} 
\int \frac{d^3 p}{(2 \pi \hbar)^3} {\rm e}^{i {\bm p} \cdot \bar{\bm r}_{12} / \hbar} \notag \\
& \ \ \ \times \underline{\cal U}_{\rm HF} ({\bm p}, {\bm r}_{12}) 
{\rm e}^{( i \hbar / 4) e {\bm B} ({\rm r}_{12}) \cdot 
(\overleftarrow{\bm \partial}_{\bm p} \times \overrightarrow{\bm \partial}_{\bm p})} \notag \\
& \ \ \ \times {\rm e}^{(i \hbar / 2) \overleftarrow{\bm \partial}_{12} \cdot \overrightarrow{\bm \partial}_{\bm p} 
- (i \hbar / 2) \overleftarrow{\bm \partial}_{\bm p} \cdot \overrightarrow{\bm \partial}_{12}} 
\underline{F} (\varepsilon_n, {\bm p}, {\bm r}_{12}), \label{L2}
\end{align}
\begin{align}
&\underline{M} ({\bm r}_1, {\bm r}_2 ; \varepsilon_n) 
\approx {\rm e}^{i I ({\bm r}_1, {\bm r}_{12}) - i I ({\bm r}_{12}, {\bm r}_2)} 
\int \frac{d^3 p}{(2 \pi \hbar)^3} {\rm e}^{i {\bm p} \cdot \bar{\bm r}_{12} / \hbar} \notag \\
& \ \ \ \times \underline{\Delta} ({\bm p}, {\bm r}_{12}) 
{\rm e}^{- ( i \hbar / 4) e {\bm B} ({\rm r}_{12}) \cdot 
(\overleftarrow{\bm \partial}_{\bm p} \times \overrightarrow{\bm \partial}_{\bm p})} \notag \\
& \ \ \ \times {\rm e}^{(i \hbar / 2) \overleftarrow{\bm \partial}_{12} \cdot \overrightarrow{\bm \partial}_{\bm p} 
- (i \hbar / 2) \overleftarrow{\bm \partial}_{\bm p} \cdot \overrightarrow{\bm \partial}_{12}} 
\underline{G}^* (\varepsilon_n, - {\bm p}, {\bm r}_{12}). \label{M2}
\end{align}

We substitute  Eqs. (\ref{J2}), (\ref{K2}), (\ref{L2}), and (\ref{M2}) into Eq. (\ref{SelfEnegyTerms}) 
to obtain the self-energy terms of the Gor'kov equation (\ref{Gor'kovEq}) in the Wigner representation. 
In addition, we expand the Hartree--Fock potential formally as 
$\underline{\cal U}_{\rm HF} ({\bm p}, {\bm r}) 
= {\cal U}_{\rm HF} ({\bm p}) \underline{\sigma}_0 + O (\underline{\Delta}^2 ({\bm p}, {\bm r}))$ 
with ${\cal U}_{\rm HF} ({\bm p})$ denoting the Hartree--Fock potential in the homogeneous normal state,  
and neglect all terms of the product of two momentum derivatives of the pair potential and Green's function 
such as $\partial \underline{\Delta} / \partial {\bm p} \times \partial \underline{G} / \partial {\bm p}$
and $\partial \underline{\Delta} / \partial {\bm p} \times \partial \underline{F} / \partial {\bm p}$. 
By following this procedure, we obtain the self-energy terms in the Wigner representation as
\begin{align}
&\int d^3 \bar{r}_{12} {\rm e}^{- i {\bm p} \cdot \bar{\bm r}_{12}} \notag \\
& \ \ \ \times \hat{\Gamma} ({\bm r}_{12}, {\bm r}_1) 
\int d^3 r_3 \hat{\cal U}_{\rm BdG} ({\bm r}_1, {\bm r}_3) \hat{G} ({\bm r}_3, {\bm r}_2; \varepsilon_n) 
\hat{\Gamma} ({\bm r}_2, {\bm r}_{12}) \notag \\
&\approx \hat{\Delta} ({\bm p}, {\bm r}_{12}) \circ \hat{G} (\varepsilon_n, {\bm p}, {\bm r}_{12}) 
+ {\cal U}_{\rm HF} ({\bm p}) \hat{\tau}_3 \circ \hat{G} (\varepsilon_n, {\bm p}, {\bm r}_{12}) \notag \\
& \ \ \ \ \ + \frac{i \hbar}{8} e {\bm B} ({\bm r}_{12}) \cdot \Bigg\{ \left( {\bm v} - \frac{\bm p}{m} \right) \notag \\
& \ \ \ \ \ \ \ \ \ \ \times \frac{\partial}{\partial {\bm p}} \left[ 3 \hat{G} (\varepsilon_n, {\bm p}, {\bm r}_{12}) 
+ \hat{\tau}_3 \hat{G} (\varepsilon_n, {\bm p}, {\bm r}_{12}) \hat{\tau}_3 \right] \Bigg\},
\label{SelfEnegyTerms3}
\end{align} 
where ${\bm v}$ is the velocity in the normal state given by 
\begin{align}
{\bm v} = \frac{\partial \varepsilon_{\bm p}}{\partial {\bm p}}, \ \ \ 
\varepsilon_{\bm p} = \frac{p^2}{2m} + {\cal U}_{\rm HF} ({\bm p}).
\end{align}

\end{document}